\documentclass[12pt]{article}

\usepackage{amsmath,amssymb,amsthm,amscd}
\usepackage[dvips]{graphics}
\usepackage{epsfig}
\usepackage{psfrag}
\makeatletter

\renewcommand\section{\@startsection{section}{1}{\z@}
                                   {-3.5ex \@plus -1ex \@minus -.2ex}
                                   {2.3ex \@plus .2ex}
                                   {\normalfont\large\bfseries}}
\renewcommand\subsection{\@startsection{subsection}{2}{\z@}
                                   {-3.25ex\@plus -1ex \@minus -.2ex}
                                   {1.5ex \@plus .2ex}
                                   {\normalfont\normalsize\bfseries}}
\renewcommand\subsubsection{\@startsection{subsubsection}{3}{\z@}
                                   {-3.25ex\@plus -1ex \@minus -.2ex}
                                   {1.5ex \@plus .2ex}
                                   {\normalfont\normalsize\bfseries}}
\renewcommand\paragraph{\@startsection{paragraph}{4}{\z@}
                                   {3.25ex \@plus1ex \@minus.2ex}
                                   {-1em}
                                   {\normalfont\normalsize\bfseries}}

\makeatother

\newdimen\tableauside\tableauside=1.0ex
\newdimen\tableaurule\tableaurule=0.4pt
\newdimen\tableaustep
\def\phantomhrule#1{\hbox{\vbox to0pt{\hrule height\tableaurule
width#1\vss}}}
\def\phantomvrule#1{\vbox{\hbox to0pt{\vrule width\tableaurule
height#1\hss}}}
\def\sqr{\vbox{%
  \phantomhrule\tableaustep

\hbox{\phantomvrule\tableaustep\kern\tableaustep\phantomvrule\tableaustep}%
  \hbox{\vbox{\phantomhrule\tableauside}\kern-\tableaurule}}}
\def\squares#1{\hbox{\count0=#1\noindent\loop\sqr
  \advance\count0 by-1 \ifnum\count0>0\repeat}}
\def\tableau#1{\vcenter{\offinterlineskip
  \tableaustep=\tableauside\advance\tableaustep by-\tableaurule
  \kern\normallineskip\hbox
    {\kern\normallineskip\vbox
      {\gettableau#1 0 }%
     \kern\normallineskip\kern\tableaurule}%
  \kern\normallineskip\kern\tableaurule}}
\def\gettableau#1 {\ifnum#1=0\let\next=\null\else
  \squares{#1}\let\next=\gettableau\fi\next}

\newcommand{\be}{\begin{equation}}
\newcommand{\ee}{\end{equation}}
\newcommand{\bea}{\begin{eqnarray}}
\newcommand{\eea}{\end{eqnarray}}
\newcommand{\ba}{\begin{array}}
\newcommand{\ea}{\end{array}}
\newcommand{\id}{\hbox{1\kern-.27em l}}

\newcommand{\CC}{\mathbb{C}}
\newcommand{\RR}{\mathbb{R}}
\newcommand{\half}{ {\textstyle \frac{1}{2}  } }
\newcommand{\ha}{\hat{a}}
\newcommand{\al}{\alpha}
\newcommand{\ga}{\gamma}
\newcommand{\Ga}{\Gamma}
\newcommand{\bet}{\beta}

\newcommand{\vka}{\varkappa}
\newcommand{\de}{\delta}
\newcommand{\vphi}{\varphi}
\newcommand{\ep}{\epsilon}
\newcommand{\vep}{\varepsilon}
\newcommand{\si}{\sigma}
\newcommand{\la}{\lambda}

\newcommand{\om}{\omega}

\newcommand{\De}{\Delta}
\newcommand{\La}{\Lambda}

\newcommand{\Ups}{\Upsilon}
\newcommand{\cN}{\mathcal{N}}
\newcommand{\cO}{\mathcal{O}}
\newcommand{\cW}{\mathcal{W}}
\newcommand{\cF}{\mathcal{F}}
\newcommand{\cT}{\mathcal{T}}

\newcommand{\cC}{{\mathcal C}}
\newcommand{\tr}{{\rm tr}}
\newcommand{\D}{{\rm d}}
\newcommand{\pa}{\partial}
\newcommand{\rar}{\rightarrow}

\newcommand{\non}{\nonumber}
\newcommand{\lb}{\langle}
\newcommand{\rb}{\rangle}
\newcommand{\SU}{\mathrm{SU}}
\newcommand{\SO}{\mathrm{SO}}

\newcommand{\U}{\mathrm{U}}

\newcommand{\rmd}{{\rm d}}
\newcommand{\rme}{{\rm e}}
\newcommand{\rmi}{{\rm i}}
\def\BR{{\mathbb R}}
\def\BC{{\mathbb C}}
\def\im{{\mathbb{I}}{\mathrm{m}}}
\def\re{{\mathbb{R}}{\mathrm{e}}}

\begin{document}

\begin{center}

\vspace*{5mm}
{\Large\sf
An $A_r$ threesome: \\
\medskip
Matrix models, $2d$ CFTs and $4d$ $\cN=2$ gauge theories
}

\vspace*{5mm}
{\large Ricardo Schiappa$^a$ and Niclas Wyllard$^b$}

\vspace*{5mm}
$^a$ CAMGSD, Departamento de Matem\'atica\\
Instituto Superior T\'ecnico\\ 
Av. Rovisco Pais 1, 1049--001 Lisboa, Portugal\\
[3mm]
$^b$ Department of Fundamental Physics\\
Chalmers University of Technology\\
S--412 96 G\"oteborg, Sweden\\
[3mm]
{\tt
schiappa@math.ist.utl.pt\,,\quad n.wyllard@gmail.com}          

\vspace*{5mm}
{\bf Abstract} 
\end{center}
\noindent
We explore the connections between three classes of theories:  $A_r$ quiver matrix models, $d=2$ conformal  $A_r$ Toda field theories and $d=4$ $\cN=2$ supersymmetric conformal  $A_r$ quiver gauge theories. In particular, we analyse the quiver matrix models recently introduced by Dijkgraaf and Vafa \cite{Dijkgraaf:2009} and make detailed comparisons with the corresponding quantities in the  Toda field theories and the $\cN=2$  quiver gauge theories. We also make a speculative proposal for how the matrix models should be modified in order for them to reproduce the instanton partition functions in quiver gauge theories in five dimensions.

\setcounter{equation}{0}
\section{Introduction}

The AGT relation \cite{Alday:2009}, which is a relation between Nekrasov partition functions \cite{Nekrasov:2002} in (conformal) $d=4$ $\cN=2$ quiver gauge theories and correlation functions in conformal field theories in two dimensions, has been studied in several papers over the past couple of months. The original conjecture \cite{Alday:2009} involves a relation between Nekrasov partition functions in $d=4$ $\SU(2)$ (or $A_1$) quiver gauge theories \cite{Gaiotto:2009} and correlation functions in the Liouville $d=2$ conformal field theory. It was subsequently extended \cite{Wyllard:2009} to a relation between the $A_r$ quiver gauge theories \cite{Gaiotto:2009} and the $d=2$ conformal $A_r$ Toda field theories. The proposals in \cite{Alday:2009, Wyllard:2009} have passed many non--trivial checks, see \textit{e.g.}~\cite{Mironov:2009b, Mironov:2009c, Marshakov:2009a}. In \cite{Alday:2009b} the $A_1$ AGT relation was extended by the inclusion of surface,  Wilson and 't~Hooft operators in the $A_1$ quiver gauge theories and proposals were made for the corresponding quantities in the Liouville theory. Another line of investigation concerns the extension to non--conformal $\SU(2)$ gauge theories \cite{Gaiotto:2009c}. A suggestion for how this approach should be modified to capture the instanton partition function for pure $\SU(2)$ gauge theory in five dimensions was presented in \cite{Awata:2009}. In~\cite{Gadde:2009}, another relation between four--dimensional and two--dimensional theories was uncovered which is similar in spirit to the AGT relation.

Recently Dijkgraaf and Vafa presented an argument explaining the AGT relations \cite{Dijkgraaf:2009}\footnote{Another argument, using M--theory, for the validity of the AGT relation was presented in \cite{Bonelli:2009}; see also the follow--up paper \cite{Alday:2009c}.}. The argument involves relating the relevant quantities in the two theories via an intermediate matrix model. The first step is to realize the gauge theories in string theory using geometric engineering \cite{Klemm:1996} and then use the relation  to matrix models via a large $N$ duality, together with the relation between matrix models and conformal field theories \cite{Kharchev:1992, Kostov:1999} to recover the AGT relation. This chain of arguments removes some of the mystery of the AGT relation. More importantly, it also implies that there are now \textit{three different ways} to compute the same quantities: using the $4d$ quiver gauge theories, using the $2d$ Toda theories, or using the $0d$ quiver matrix models. In all three cases a Riemann surface plays a crucial role: in the gauge theory the Riemann surface is related to the Seiberg--Witten curve, in the Toda theory the Riemnan surface is the manifold on which the theory is defined, and in the matrix model the Riemann surface is the spectral curve arising from the loop equations in the large $N$ limit.

The goal of this paper is to develop and exemplify how calculations are performed in the matrix model framework. We will rederive several known results in quiver gauge theories and Toda field theories from the matrix model integrals. We also make a speculative proposal for how the matrix models should be modified to reproduce the Nekrasov partition function for quiver gauge theories in five dimensions.

In the next section we review the AGT relation for the case of the $A_r$ theories and in section \ref{dv} we describe the $A_r$ quiver matrix models introduced in \cite{Dijkgraaf:2009}. In the two subsequent sections we then perform several matrix model calculations and compare the results with the Toda theories and the quiver gauge theories: In section \ref{sA1} we treat the $A_1$ model and in section \ref{sAr} we discuss the $A_r$ models for general $r$. Finally, in section \ref{sq} we describe our proposal for how the matrix models should be modified in order  to describe the Nekrasov partition function for quiver gauge theories in five dimensions. In the appendix some technical details are collected.

\medskip
\noindent {\bf Note added:} After this paper was finished \cite{Itoyama:2009,Eguchi:2009} appeared which have some overlap with some parts of this paper.

\setcounter{equation}{0}
\section{The $A_r$ AGT relation}\label{agt}

In this section we review the AGT proposal for the class of theories based on the $A_r$ Lie algebras. We start with a brief recap of the $A_r$ Toda field theories, followed by a summary of the $A_r$ quiver gauge theories and then describe the AGT relation connecting the two classes of theories.

\subsection{The $A_r$ Toda field theories} \label{sToda}

The $A_{r}$ Toda field theories are defined by the action
\be \label{tact}
S = \int \D^2\si \sqrt{g}\left[ \frac{1}{8\pi} g^{ad}\lb \pa_{a}\phi, \pa_d \phi \rb + \mu \sum_{i=1}^{N-1} \rme^{b\lb e_i,\phi\rb} + \frac{\lb Q\rho, \phi \rb}{4\pi}R \,\phi \right],
\ee 
where $g_{ad}$ ($a,d=1,2$) is the metric on the two--dimensional worldsheet,  and $R$ is the worldsheet curvature. The $e_i$ are the simple roots of the $A_{r}$ Lie algebra, $\lb \cdot,\cdot\rb$ denotes the scalar product on the root space, $\rho$ is the Weyl vector (half the sum of all positive roots) and the $r$--dimensional vector of fields $\phi$ can be expanded as $\phi = \sum_i \phi_i e_i$. The $A_r$ Toda theory is conformal provided $Q$ and $b$ are related via
\be \label{Q}
Q = \left( b+\frac{1}{b} \right) \,.
\ee
The central charge is \cite{Mansfield:1982}
\be \label{ct}
c = r + 12 \,Q^2 \,\lb \rho, \rho\rb = r \left( 1+ \left(r+1\right)\left(r+2\right)\left(b+\frac{1}{b}\right)^2 \right).
\ee

The general form of a three--point correlation function in a 2$d$ conformal field theory is \cite{Belavin:1984}
\be \label{3pt}
\lb V_{\al_1}(z_1,\bar{z}_1) V_{\al_2}(z_2,\bar{z}_2)V_{\al_3}(z_3,\bar{z}_3)\rb=\frac{C(\al_1,\al_2,\al_3)}{ |z_{12}|^{2(\De_1+\De_2-\De_3)} |z_{13}|^{2(\De_1+\De_3-\De_2)} |z_{23}|^{2(\De_2+\De_3-\De_1)}} \,.
\ee

The Liouville theory is identical to the $A_1$ Toda field theory and has a set of primary fields 
\be
V_{\al} = \rme^{2 \al \phi }\,.
\ee
The correlation function of three primary fields in the Liouville theory is \cite{Dorn:1994} (see also \cite{Teschner:2003})
\bea \label{Lthree}
&&C(\al_1,\al_2,\al_3) =  \left[\pi\mu\ga(b^2)b^{2-2b^2}\right]^{\frac{Q-\al_1-\al_2-\al_3}{b}} \times \\ &&\times\,
    \frac{ \Ups(b) \Ups(2\al_1)
     \Ups(2\al_2)  \Ups(2\al_3) }{\Ups(\al_1+\al_2+\al_3 -Q) \Ups(-\al_1+\al_2+\al_3)
\Ups(\al_1-\al_2+\al_3) \Ups(\al_1+\al_2-\al_3)}\,, \non
\eea
with $\ga(b^2) = \frac{\Ga(b^2)}{\Ga(1-b^2)}$ and  $\Ups(x) = 1/[\Ga_2(x|b,b^{-1}) \Ga_2(Q-x|b,b^{-1})]$ where $\Ga_2(z| \ep_1,\ep_2 )$ is the Barnes double Gamma function \cite{Barnes:1901}.

In the Toda theories with $r>1$ ($\cW$) primary fields can be defined in analogy with the Liouville case via 
\be \label{prim}
V_{\alpha}=\rme^{\lb\alpha,\phi\rb}\,.
\ee

Recently it was shown \cite{Fateev:2005, Fateev:2007b} that in the special case when one of the $\al$'s takes one of the two special values 
\be \label{spec}
\chi = \vka \La_{1} \qquad \mathrm{or} \qquad \chi = \vka \La_{r}\,,
\ee
where $ \La_{1}$ ($\La_{r}$) is the highest weight of the fundamental (anti--fundamental) representation of the $A_{r}$ Lie algebra and $\vka$ is a complex number, the three--point function is given by  (\ref{3pt}) with
\bea \label{t3pt}
C(\al_1,\al_2,\chi)& =  & \left[\pi\mu\ga(b^2)b^{2-2b^2}\right]^{\frac{\lb 2Q\rho-\al_1-\al_2-\chi ,\rho \rb}{b}} \times \\ &
   \times&\!\!\! \frac{\left(\Ups(b)\right)^{r}\Ups(\vka)\prod_{e>0}\Ups\big(\lb Q\rho-\al_1,e\rb\big)
    \Ups\big(\lb Q\rho-\al_2,e\rb\big)}{\prod_{ij}\Ups\big(\frac{\vka}{r+1}+
    \lb\al_1-Q\rho,h_i\rb+\lb\al_2-Q\rho,h_j \rb\big)}\,, \non
\eea
where the product in the numerator is over all positive roots and in the denominator the $h_i$ are the weights of the representation with highest weight $\La_{1}$, \textit{cf.}~(\ref{hs}). (The result for $\chi=\vka \La_r$ is obtained by replacing $h_i$ by $h'_i = -h_{r+2-i}$.) 

Higher--point correlation functions in any CFT can be related to the three--point function of primary fields, which therefore determines the entire theory \cite{Belavin:1984}. Note that when $r>1$ knowledge of the three--point function of $\cW$ primary fields (\ref{t3pt}) does not determine all higher--point correlation functions, see \textit{e.g.}~\cite{Bowcock:1993,Wyllard:2009}.

As an example, consider a four--point function. It is convenient to fix three points to $0, 1, \infty$ and use a bra--ket notation which has the property $\lb \al|\al \rb=1$, and is such that
\be
\lb \al_1| V_{\al_2}(1) V_{\al_3}(z) |\al_4 \rb = \lb V_{Q-\al_1}(0)  V_{\al_2}(1) V_{\al_3}(z) V_{\al_4}(\infty) \rb\,.
\ee
Inserting a complete set of states we find
\be \label{cb}
\lb \al_1| V_{\al_2}(1) V_{\al_3}(z) |\al_4 \rb = \int \D \si \sum_{{\bf k},{\bf k}'} 
 \frac{  \lb \al_1| V_{\al_2}(1) |\psi_{\bf k}(\si) \rb   \lb \psi_{{\bf k}'}(\si)| V_{\al_3}(z) |\al_4 \rb }{\lb \psi_{-\bf k}(\si) |\psi_{-\bf k'} (\si)\rb} \,.
\ee
Here the intermediate states $|\psi_{\bf k}(\si)\rb$ are descendants of the primary state labelled by $\si$. (Throughout this paper we will label the internal momenta by $\si$ reserving the symbol $\al$ for the external momenta.)

In the Liouville case it can be shown that $\lb \psi_{{\bf k}'}(\si)| V_{\al_3}(z) |\al_4 \rb$ is proportional to $\lb \si| V_{\al_3}(z) | \al_4 \rb$ \cite{Belavin:1984} and  hence (\ref{cb}) can be calculated perturbatively. The ratio
\be
\frac{\sum_{{\bf k},{\bf k}'}  \lb \al_1| V_{\al_2}(1) |\psi_{\bf k}(\si) \rb \lb \psi_{-\bf k}(\si) |\psi_{-\bf k'} (\si)\rb^{-1} \lb \psi_{{\bf k}'}(\si)| V_{\al_3}(z) |\al_4 \rb }{ \lb \al_1| V_{\al_2}(1) |\si \rb \lb \si| V_{\al_3}(z) |\al_4 \rb }
\ee
is called a conformal block. General $n$--point functions can be dealt with in an analogous manner. They depend on $(n-3)$ cross ratios. 

In the $r>1$ case the situation is a little more involved, see \cite{Wyllard:2009} for a discussion.

\subsection{The $A_r$ quiver gauge theories and Nekrasov partition functions} \label{squiver}

In \cite{Gaiotto:2009} a class of conformal 4$d$ $\cN=2$ generalised  $A_r$  quiver gauge theories were introduced. This class of theories was denoted $\cT_{(n,g)}(A_{r})$. The simplest example in this class of theories is the theory  with a single $\SU(r+1)$ gauge factor with  $2(r+1)$ matter hypermultiplets in the fundamental representation of the gauge group. But the $\cT_{(n,g)}(A_{r})$  class of theories includes many more theories, not all of which are conventional weakly--coupled gauge theories. The $\cT_{(n,g)}(A_{r})$  theories can be viewed as arising from the six--dimensional $A_{r}$ (2,0) theory \cite{Witten:1995} compactified on $C\times \RR^4$ where $C$ is a genus $g$ Riemann surface with $n$ punctures. The genus of the Riemann surface depends on the number of loops in the (generalised) quiver diagram. The punctures are due to  codimension 2 defects filling $\RR^4$ and intersecting $C$ at points, and were argued in \cite{Gaiotto:2009} to be classified by partitions of $r+1$ (which can be represented graphically in terms of Young tableaux). One can therefore associate a Young tableau to each puncture. In the case of the $A_1$ theories there is only one kind of non--trivial puncture. In the above example (a $\cT_{4,0}(A_r)$ theory) there are two kinds of punctures. These are associated with the  factors in the $ \U(1)^2\SU(r+1)^2$ subgroup of the flavour symmetry group.  The punctures associated with the $\SU(r+1)$ factors are called full punctures and involve $r$ mass parameters each and the punctures associated with the $\U(1)$ factors are called basic punctures and involve one mass parameter each. We refer to \cite{Gaiotto:2009} for further details. 

A fundamental object in an $\cN=2$ gauge theory is the Nekrasov partition function (from which the prepotential can be obtained). The partition function factorises into two parts as  
\be \label{z}
Z = Z_{\rm pert}\,Z_{\rm inst}\,,
\ee 
where $Z_{\rm pert}$ is the contribution from perturbative calculations (because of supersymmetry there are contributions only at tree and one--loop level), and $Z_{\rm inst}$ is the contribution from instantons. 
 The most efficient method to obtain $Z_{\rm inst}$ is via the instanton counting method of Nekrasov~\cite{Nekrasov:2002}. This approach involves deforming the $\cN=2$ gauge theory with two parameters $\ep_1$ and $\ep_2$ which belong to an  $\SO(2){\times}\SO(2)$ subgroup of the $\SO(4)$ Lorentz symmetry.  We should stress that one needs the theory to be weakly coupled to be able to apply the instanton counting method.  

As an example, the instanton partition function in the $\SU(r+1)$ theory with $2(r+1)$ fundamentals can be written \cite{Nekrasov:2002} (see also~\cite{Flume:2002})
\be \label{Zinst}
Z_{\rm inst} = \sum_{\vec{Y}} y^{|\vec{Y}|}  \prod_{m,n=1}^{r+1} \prod_{s\in Y_m} \frac{P(\ha_m,Y_m,s)}{E(\ha_m-\ha_n,Y_m,Y_n,s)(E(\ha_m-\ha_n,Y_m,Y_n,s) -\ep) }\,,
\ee
where $y=\rme^{2\pi i \tau}$ and the sum is over the  ($r+1$)-dimensional vector of Young tableaux, $\vec{Y} = (Y_1,Y_2,\ldots,Y_{r+1})$ and $|\vec{Y}|$ (the instanton number) is the total number of boxes in all the $Y_m$'s. The $\ha_m$  parameterise the Coulomb branch of the theory and satisfy $\sum_{i=1}^{r+1} \ha_i =0$. It is convenient to write $\ha = \sum_{i=1}^{r} a_i \,e_i$ where $e_i$ are the simple roots of the $A_{r}$ Lie algebra. In the particular case of $\SU(2)$ this translates into $\ha = (a,-a)$. In (\ref{Zinst})  $\ep \equiv \ep_1+\ep_2$ and 
\be
E(x,Y_m,Y_n,s) = x-\ep_1 L_{Y_n}(s) + \ep_2(A_{Y_m}(s)+1)\,,
\ee
 where $s=(i,j)$ and $i$ refers to the vertical position and $j$ to the horizontal position of the box. Furthermore, $L_{Y_n} = k_{n,i}-j$ and $A_{Y_m} = k^T_{m,j} - i$, where $k_{n,i}$ is the length of the $i$th row of $Y_n$ and $k^T_{m,j}$ is the height of the $j$th column of $Y_m$. Finally,  
\be
P(x,Y_i,s) = \prod_{f=1}^{2r+2} (x-(j-1)\ep_1-(i-1)\ep_2 - m_f)\,,
\ee
where the $m_f$ are the masses of the matter fields (suitably defined). 

The perturbative (one--loop) piece in (\ref{z}), $Z_{\rm pert}$, is a product of various factors. For $\SU(r+1)$ the gauge field contributes a factor
\be
\prod_{i<j}^{r} \frac{1}{\Ga_2(\ha_i-\ha_j-\ep_2|\ep_1,\ep_2)\Ga_2(\ha_i-\ha_j-\ep_1|\ep_1,\ep_2)}\,,
\ee
where $\Ga_2(x|\ep_1,\ep_2)$ is the Barnes double gamma function \cite{Barnes:1901}, and each of the massive hypermultiplets transforming in the fundamental representation of the gauge group contributes a factor
\be \label{gapert}
\prod_{i=1}^{r} \Ga_{2}(a_i-m_f+\ep|\ep_1,\ep_2)\,.
\ee

\subsection{The AGT relation}

The AGT relation is a relation between the two classes of theories discussed in sections \ref{sToda} and \ref{squiver}. Up to an overall factor, the instanton partition function of a $\cT_{(n,g)}(A_r)$ theory is (conjectured to be) equal to a chiral block of an $n$--point correlation function in the $A_r$ Toda field theory formulated on a genus $g$ surface. In other words, the punctures correspond to insertions of vertex operators in the Toda theory. For example, the instanton partition function in the $\SU(r+1)$ theory with $2r+2$ fundamentals is equal to the chiral block (in a specific channel) of the four--point function in the $A_r$ Toda theory on the sphere. The momenta of the vertex operators, $\al_i$, are mapped to the masses,  $m_i$, in the gauge theory. This relation is linear (the exact form depends on conventions for the gauge theory masses). Furthermore, the internal momenta in the chiral block, $\si_k$, are linearly related to the $a_k$ Coulomb moduli. Finally, the parameters $\ep_1$ and $\ep_2$ in the instanton partition function are related to the parameter $b$ in the Toda theory via,
\be
b = \sqrt{\frac{\ep_1}{\ep_2}} \,,\qquad \frac{1}{b} = \sqrt{\frac{\ep_2}{\ep_1}}\,.
\ee 
The most common choice is to set $\ep_1=b$, $\ep_2=\frac{1}{b}$. There is also a slight extension of the above result where the full partition function including the perturbative piece is related to the full correlation function including the three--point pieces. For further details about the AGT relation, see~\cite{Alday:2009,Wyllard:2009}.

\setcounter{equation}{0}
\section{The matrix model approach of Dijkgraaf and Vafa}\label{dv}

The first step in the analysis of \cite{Dijkgraaf:2009} is to realise  the relevant quiver gauge theories in string theory using geometric engineering \cite{Klemm:1996} and then use the results in~\cite{Dijkgraaf:2002a} to relate the topological string partition function (which is equal to the Nekrasov partition function) to a matrix model calculation. This argument works provided $\ep_1=-\ep_2 = g_s$. How to deal with the case of general $\ep_{1,2}$ was also proposed in \cite{Dijkgraaf:2009} and will be discussed later in this section. 

In the case of interest to us the relevant geometries are $A_{r}$ singularities
\be
uv-x^{r+1}=0.
\ee
(There is also a decoupled $\CC$ factor parametrised by $z$.)
The matrix model corresponding to this geometry is the so called $A_{r}$ quiver matrix model \cite{ Kharchev:1992,Kostov:1992}. It involves $r$ hermitian $N_i\times N_i$ matrices $\Phi_i$ ($i=1,\ldots,r$) as well as the $N_i\times N_j$ matrices $B_{ij}$. The matrix model partition function is (proportional to)
\be \label{ade}
\int \prod_i \D\Phi_i \prod_{i,j} \D B_{ij} \exp\left[ - \frac{1}{g_s}\prod_{i<j} \tr \left[ (2\de_{ij}-A_{ij})( B_{ij} \Phi_j B_{ji} - B_{ji} \Phi_i B_{ij}) \right] \right]\,,  
\ee
where $A_{ij}$ is the Cartan matrix of the $A_r$ Lie algebra, \textit{i.e.}, $A_{ij}=\lb e_i,e_j\rb$. In terms of the eigenvalues of $\Phi_i$, $\la_i^I$ ($I=1,\ldots,N_i$), the partition function (\ref{ade}) becomes, after integrating out the $B_{ij}$,
\be
\int \prod_{iI} \D \la^I_i \prod_{(i,I)<(j,J)} |\la^I_i-\la_j^J|^{A_{ij}} .
\ee
Here $(i,I)<(j,J)$ if $i<j$ or $i=j$, $I<J$.

There is a close relationship between the $A_{r}$ matrix models and conformal field theory \cite{Kharchev:1992, Kostov:1999}. In this relation $r$ free bosons, $\vphi_i$, in a $2d$ conformal field theory are related to matrix model quantities in the following way\footnote{ This relation holds when the matrix model potential is zero which is the case we are interested in. The factor of $g_s$ is unconventional but convenient for our purposes. }
\be \label{cftom}
\pa \vphi_i(z)  = \frac{1}{g_s} \tr\left( \frac{1}{z-\Phi_i}\right)\,.
\ee 
 Because of the relation (\ref{cftom}) the free boson CFT vertex operator $\hat{V}_\al(z) = \rme^{\al^i \vphi_i(z)}$ translates into
\be 
\prod_i \det(z -\Phi_i)^{\al_i/g_s}\,.
\ee
Here $\al^i = \lb e_i, \al\rb$ where $\al = \sum_i \al^i \La_i$ (see appendix \ref{Alie} for a summary of the Lie algebra terminology). Correlation functions of (free boson) CFT vertex operators can therefore be calculated using matrix model technology. The correlation function\footnote{In this paper we restrict our attention to correlation functions on genus 0 surfaces.}  
\be \label{VV}
\lb \hat{V}_{\al_1/g_s}(z_1) \cdots \hat{V}_{\al_k/g_s}(z_k)\rb_{\al_0,N}
\ee
translates into 
\bea \label{mmVV}
\int  \prod_i \D\Phi_i  \prod_{i,j} \D B_{ij} \prod_i \det(z_1-\Phi_i)^{\al^i_1/g_s} \cdots \prod_i \det(z_k-\Phi_i)^{\al_k^i/g_s} \non \\ \exp\left[ - \frac{1}{g_s}\prod_{i<j} \tr \left ( (\de_{ij}-A_{ij})( B^{ij} \Phi^J \tilde{B}^{ji} - B^{ji} \Phi^i \tilde{B}^{ij}) \right)\right]  .
\eea
In (\ref{VV}) the subscript $\al_0$ refers to the fact that there is an extra $\al_0$ charge at infinity. This means that the correlation function really involves $k+1$ vertex operators. Furthermore, in our conventions,
\be \label{AN}
A_{ij} N_j = \al_0^i/g_s - \sum_{n=1}^k \al^i_n/g_s \,.
\ee 
To analyze the matrix model expression (\ref{mmVV}) one can use the following relation $\det(z-\Phi)^{\al/g_s} = \exp(\frac{\al}{g_s} \, \tr\, \log(z-\Phi)\,)$. The insertions therefore effectively induce the matrix model potentials (of multi Penner type)
\be \label{mmpot}
W_i(\Phi_i) = \tr\, \sum_{a=1}^k \al_a^i \log(z_a-\Phi_i)\,,
\ee
and the matrix models can therefore be analyzed using standard techniques.

In terms of the eigenvalues the matrix model correlation function is 
\be \label{mmeig}
\int \prod_{iI} \D \la^I_i \prod_{(i,I)<(j,J)} |\la^I_i-\la_j^J|^{A_{ij}}\prod_{i,I} (z_1-\la^I_i)^{\al^i_1/g_s} \cdots \prod_{i,I} (z_k-\la_i^I)^{\al_k^I/g_s} .
\ee
In this expression we have left the integration contour unspecified. The choice of integration contour turns out to be quite subtle and will be discussed in later sections. In previous applications of matrix models to supersymmetric gauge theories \cite{Dijkgraaf:2002a} the matrix models should properly be thought of as holomorphic matrix models with a choice of contour, see, \textit{e.g.}, \cite{Lazaroiu:2003} for a discussion. In the present case, there are also additional subtleties  since the  potentials are logarithmic. 

The proposal of Dijkgraaf and Vafa \cite{Dijkgraaf:2009} is as follows: to connect the chiral correlation functions of the vertex operators $\hat{V}$ involving the free scalar fields $\vphi_i$ to the correlation functions of the chiral vertex operators $V$ involving the fields $\phi_i$ in the $A_r$ Toda field theory one should take the large $N$ limit and identify the $\al_a$'a (including $\al_0$) with the external momenta of the Toda theory vertex operators. Furthermore, the matrix model potential (\ref{mmpot}) has stationary points which in the large $N$ limit expand into cuts. The corresponding filling fractions $g_s N_i^m$ (subject to the constraint  $\sum_{m=1}^{k-2} N^m_i = N_i$) are related to the internal momenta $\si_m$ in the Toda theory ($m$ label the internal momenta and $i$ label the components of each of the $\si_m$).

This proposal shares many similarities with the earlier work \cite{Dijkgraaf:2002a}, but we should stress that in \cite{Dijkgraaf:2009} the number of terms in the potential is related to the number of nodes in the gauge theory quiver, whereas the number of matrices is related to the rank of the gauge group; in \cite{Dijkgraaf:2002a} the roles were reversed. 

In \cite{Dijkgraaf:2009} there is also a discussion of the AGT relation using brane probes; this approach will not be used in this paper.

The analysis so far only involves $g_s$, \textit{i.e.}, $\ep_1=-\ep_2$. As mentioned above there is a further refinement of the matrix model that is needed to treat the case with general $\ep_{1,2}$. In \cite{Dijkgraaf:2009} it was suggested that the required modification is the so called $\beta$ deformation (or  $\beta$ ensemble) \cite{Mehta:2004}. This deformation changes (\ref{mmeig}) to  
\be \label{mmbeta}
\int \prod_{iI} \D \la^I_i \prod_{(i,I)<(j,J)} |\la^I_i-\la_j^J|^{\beta A_{ij}}\prod_{i,I} (z_1-\la^I_i)^{\sqrt{\bet} \al^i_1/g_s} \cdots \prod_{i,I} ( z_k-\la_i^I)^{\sqrt{\bet} \al_k^i/g_s} 
\ee
with the identification $\bet=-\ep_2/\ep_1$ and $g_s = \sqrt{-\ep_1\ep_2}$. Also, (\ref{AN}) changes to
\be
\bet A_{ij} N_j = \frac{\sqrt{\bet}}{g_s} \al_0^i -\frac{\sqrt{\bet}}{g_s} \sum_{n=1}^k \al^i_n \,.
\ee 
We should stress that for general $\bet$ the above integral (\ref{mmbeta}) can no longer be viewed as arising from an integral over matrices in any reasonable way. Therefore, strictly speaking, we are no longer dealing with a matrix model. Sometimes the model for general $\bet$ is called a generalised matrix model, but we will by a slight abuse of terminology continue to call it a matrix model. 

In the next section we analyze  various aspects of the above matrix model for the case of the $A_1$ theory and make detailed calculations and comparisons with the corresponding  expressions in the $4d$ $A_1$ quiver gauge theories  and the $2d$ Liouville theory. In section \ref{sAr} a similar analysis will be performed for the $A_r$ theories with $r>1$.

\setcounter{equation}{0}
\section{The $A_1$ matrix model}\label{sA1}

In this section we perform several calculations in the $A_1$ matrix model. The resulting expressions are compared to the corresponding expressions in the Liouville theory and the $A_1$ quiver gauge theories.

\subsection{The three--point function}\label{sA13pt}

Our first example is the matrix model three--point function:
\be \label{A1mm3}
\frac{1}{(2\pi)^NN!} \int \prod_{I} \D \la^I \prod_{I<J} |\la^I-\la^J|^{2\beta} \prod_{I} (\la^I)^{ 2\al_1/\vep_1} (1-\la^I)^{2 \al_2/\vep_1} \,.
\ee
Note that $1/\vep_1 = \sqrt{\bet}/g_s$\footnote{When referring to matrix model quantities we will use the notation $\vep_i$ rather than $\ep_i$ since it will turn out that our conventions are such that $\vep_i=-\ep_i$.}. This complicated looking integral is the so--called Selberg integral \cite{Selberg:1944} which can be evaluated exactly with the result\footnote{ In appendix \ref{ortho} we present an alternative derivation of this result when $\bet=1$ using orthogonal polynomials. }  
\be  \label{A1sel}
\frac{1}{(2\pi)^N} \prod_{I=1}^{N} \frac{\Ga(2\al_1/\vep_1+1+(I-1)\bet)\Ga(2\al_2/\vep_1+1+(I-1)\bet) \Ga(I \bet) }{\Ga(2\al_1/\vep_1+2\al_2/\vep_1+2+(I+N-2)\bet)\Ga(\bet)} \,.
\ee
In evaluating the above integral we assumed that the choice of integration contour is such that the $\la^I$'s are integrated over the interval $[0,1]$. (It is possible to perform changes of variables in the above integral to obtain other integration ranges; see \textit{e.g.}~\cite[section 17.5]{Mehta:2004}.) 
Using the result
\be
\Ga(z) = \sqrt{2\pi}(-\vep_1)^{1/2-z}\frac{\Ga_{2}(-z \vep_1 |-\vep_1,-\vep_2)}{\Ga_{2}(-z \vep_1 - \vep_2 |-\vep_1,-\vep_2)} \,,
\ee
where $\Ga_2(x|-\vep_1,-\vep_2)$ is the Barnes double gamma function \cite{Barnes:1901}, together with $\bet=-\frac{\vep_2}{\vep_1}$ and 
\be \label{Gaid}
\prod_{I=1}^N \Ga(z+(I-1)\bet) = (2\pi)^{\frac{N}{2}}  (-\vep_1)^{\frac{N}{2} - N z} \frac{\Ga_{2}(-z \vep_1+(N-1)\vep_2 |-\vep_1,-\vep_2)}{\Ga_{2}(-z \vep_1 - \vep_2 |-\vep_1,-\vep_2)} \,,
\ee
we find that (\ref{A1sel}) equals (here $\Ga_2(x)$ is short--hand for $\Ga_2(x|-\vep_1,-\vep_2)$ and $\vep\equiv \vep_1+\vep_2$)
\bea
&&\left (\frac{ \vep_1^{N\bet -2\bet+1}}{\Ga(\bet)} \right)^N   \times \\   
&&\frac{\Ga_{2}(-2\al_1 + N\vep_2-\vep )\Ga_{2}(-2\al_2 +N\vep_2-\vep )\Ga_{2}(N\vep_2) \Ga_{2}(-2\al_1-2\al_2 + N\vep_2 - 2\vep )}{\Ga_{2}(-2\al_1  - \vep )\Ga_{2}(-2\al_2 - \vep )\Ga_{2}( 0 ) \Ga_{2}(-2\al_1- 2 \al_2 +2N\vep_2-2\vep ) }. \non
\eea
Finally, using $N\vep_2 = -N\beta\vep_1 = (-\al_0+\al_1+\al_2)$ we obtain
\bea \label{A13pt}
&&\left (\frac{\vep_1^{N\bet-2\bet+1}}{\Ga(\bet)} \right)^N \times \\   
&& \frac{\Ga_{2}(-\al_0{-}\al_1{+}\al_2{-}\vep )\Ga_{2}(-\al_0{+}\al_1{-}\al_2{-}\vep )\Ga_{2}({-}\al_0{+}\al_1{+}\al_2 ) \Ga_{2}(-\al_0{-}\al_1{-}\al_2{-}2\vep )}{\Ga_{2}(-2\al_1  - \vep )\Ga_{2}(-2\al_2 - \vep )\Ga_{2}(0) \Ga_{2}(-2\al_0 -2\vep ) }. \non
\eea
In general, the three--point function in a $2d$ conformal field theory does not factorise into holomorphic and anti--holomorphic parts so there is no unambiguous meaning to a `chiral three--point function'.  However, after suitably rescaling the vertex operators  with multiplicative factors depending on their momenta, the Liouville three--point function (\ref{Lthree})  can be written as (recall that $\al_i^* = Q-\al_i$)
\bea \label{chiralL}
  &&   \!\!  \!\!  [\Ups(\al_1+\al_2+\al_3 -Q) \Ups(-\al_1+\al_2+\al_3)
\Ups(\al_1-\al_2+\al_3) \Ups(\al_1+\al_2-\al_3)]^{-1} = \non \\ 
&&   \!\!  \!\!  |\Ga_b(2Q {-} \al_1{-}\al_2{-}\al_3) \Ga_b(Q{+}\al_1{-}\al_2{-}\al_3)\Ga_2(Q{-}\al_1{+}\al_2{-}\al_3) \Ga_2(\al_1{+}\al_2{-}\al_3) |^2\,,
\eea 
where $\Ga_b(x)$ is short--hand for $\Ga_2(x |b,b^{-1})$. Therefore, for the Liouville theory there is a natural definition of a chiral three--point function as the ``square root" of (\ref{chiralL}). After a suitable redefinition of the matrix model vertex operators, we see that the matrix model three--point function (\ref{A13pt}) precisely captures the chiral part of the Liouville three--point function, provided that $\al_0 \rightarrow \al_3$ and we identify 
\be \label{epb}
\vep_1=-b  \,,\quad \vep_2=-1/b\,.
\ee
The matrix model expression can also be compared with (the perturbative part of) the Nekrasov partition function of the corresponding gauge theory,  which in the present case is the so called $T_2$ (or $\cT_{3,0}(A_1)$) theory --- a theory of four free hypermultiplets \cite{Gaiotto:2009}. Redefining the matrix model vertex operators as above  we are left with the four $\Ga_2$ factors in the numerator of (\ref{A13pt}). These are of precisely the right form to reproduce (the perturbative part of) the $T_2$ theory, \textit{cf.}~(\ref{gapert}) (with a suitable definition of the four masses).  (Possibly one can also make sense of (\ref{A13pt}) within the framework in \cite{Ponsot:2001}.)

We should also mention that the derivation in  \cite{Dorn:1994} of the full three--point function in the Liouville theory was based on an argument which involved a complex version of the above integral. Although the equations are similar the logic was different. 

\subsection{Higher--point functions} \label{A1high}

One can also consider higher--point functions in the matrix model. A tractable example is a four--point function, where one of the $\al_i$, $\al_3$ say, is equal to $\vep_1/2$ or $\vep_2/2$,  \textit{i.e.}~the integral
\be \label{L4pt}
\int \prod_{I} \D \la^I \prod_{I<J} |\la^I-\la^J|^{2\beta} \prod_{I} (z-\la^I)^{2\al_3/\vep_1} (\la^I)^{ 2\al_1/\vep_1} (1-\la^I)^{2\al_2/\vep_1} \,,
\ee
with $\al_3$ equal to $\vep_1/2$ or $\vep_2/2$. In \cite{Kaneko:1993} it was shown that the above integral satisfies the hypergeometric differential equation:
\be \label{hyper}
 z\, (1-z)\,\frac{\D^2 F(z)}{\D z^2} + [C - (A+B+1)\, z] \, \frac{\D F(z)}{\D z} - A\, B\,  F(z) =0\,,
\ee
where, if $\al_3=\vep_1/2$,  
\be \label{ABC1}
A=-N\,, \quad B=\frac{1}{\bet}(2\al_1/\vep_1 +2\al_2/\vep_1 + 2) + N-1\,,\quad   C=\frac{1}{\bet}(2\al_1/\vep_1 + 1)\,,
\ee
and, if  $\al_3=\vep_2/2$, 
\be  \label{ABC3}
A=\bet N\,, \quad B=-(2\al_1/\vep_1 +2\al_2/\vep_1 + 1) + \bet(2-N)\,, \quad C=-2\al_1/\vep_1 + \bet\,.
\ee 

On the other hand, it is known that in the Liouville theory the four--point function 
\be \label{bVVV}
\lb V_{-b/2}(z) V_{\al_1}(0) V_{\al_2} (1) V_{\al_0}(\infty) \rb\,,
\ee
satisfies the equation 
\bea
&&\Big[ \pa^2_z  -b^2\frac{2z-1}{z(z-1)} \pa_z  +b^2\frac{\De(\al_1)}{z^2}  +b^2\frac{\De(\al_2)}{(z-1)^2} \non \\ &&\; - b^2 \frac{\De(-b/2)+\De(\al_1)+\De(\al_2)-\De(\al_0) }{z(z-1)}  \Big] H(z)=0\,,
\eea
where $\De(\al) = \al(Q-\al)$. 
After writing $H(z) = z^{b\al_1}(1-z)^{b\al_2}F(z)$ the above equation reduces to the hypergeometric equation (\ref{hyper}) with 
\be \label{ABC2}
A = b(\al_1+\al_2-\al_0-b/2)\,,\quad  B=b(\al_1+\al_2+\al_0 -\frac{3}{2}b-\frac{1}{b})\,, \quad C= b(2\al_1 - b) \,.
\ee
Using $\bet = -\frac{\vep_2}{\vep_1}$ and $-\vep_2 N -\al_0+\al_1+\al_2+\vep_1/2 = 0$ (compared to the corresponding expression for the three--point function there is now an extra term coming from $\al_3=\vep_1/2$), we see that (\ref{ABC1}) and (\ref{ABC2}) agree provided we use the identifications (\ref{epb}). 
This analysis shows that the matrix model integral (\ref{L4pt}) with the above identifications is proportional to the chiral block in the Liouville CFT.  
Note that for the special correlation function (\ref{bVVV}) the internal momentum is restricted to two discrete values corresponding to the two solutions to the hypergeometric equation. 
The case with an insertion of the vertex operator $V_{-\frac{1}{b}}(z)$ can be treated analogously and compared to (\ref{ABC3}). 
The choice of signs in (\ref{epb}) differs from the usual convention and means that $\vep_i=-\ep_i$, but note that both  $\ep_1 \ep_2$ and $\frac{\ep_2}{\ep_1}$ are unaffected. We could make sign changes elsewhere to restore the usual rule but this would clutter some of the above formul\ae.  

One can also check that the above expression agrees with the corresponding Nekrasov partition function. This was anticipated in \cite{Wyllard:2009} and derived in detail in~\cite{Mironov:2009b}. We briefly recall the argument here. The instanton partition function is as in (\ref{zinst}) with $m,n=1,2$ and $(\ha_1,\ha_2)=(a,-a)$. If we tune the Coulomb modulus $a$ to fulfill $P(a) = 0$ by setting $a=m_1$, then only terms in the sum with only $Y_2$ non--empty give non--vanishing contributions. If we furthermore set $m_2=-a-\ep_1$ then only those $Y_2$ tableaux that have boxes only in the first column survive (in other words $k_{2,j}^T$ is only non--zero for $j=1$, so that $i=1,\ldots,k_{2,1}^T$ and $k_{2,i}=1$). Next using the AGT relation in the form
\be \label{mal}
m_1 = -\frac{\ep}{2}+\al_1+\al_3\,,\; m_2 = \frac{\ep}{2}-\al_1+\al_3\,,\; m_3 = -\frac{\ep}{2}+\al_4+\al_2\,,\; m_4 = \frac{\ep}{2}-\al_4+\al_2\,,
\ee
one finds $\al_3=-\ep_1/2$ and 
\be \label{zinst}
Z_{\rm inst} = \sum_{l=0}^{\infty} \frac{(A)\ell(B)_{\ell}}{(C)_{\ell} }\frac{y^\ell}{\ell!} \,,
\ee
where  $(X)_n= X (X+1) \cdots (X+n-1)$ is the Pochhammer symbol and 
\be  \label{ABC4}
A=  (\al_1{+}\al_2{-}\al_4{-}\frac{\ep_1}{2} )/\ep_2 \,, \quad B= (\al_1{+}\al_2{+}\al_4{-}\frac{3}{2}\ep_1{-}\ep_2 )/\ep_2\,, \quad C=-(2\al_1{-}\ep_1)/\ep_2 \,,
\ee 
which agrees with (\ref{ABC2}) provided $\al_4\rar\al_0$, $\ep_1=b$ and $\ep_2=\frac{1}{b}$. Together with the fact that the expression (\ref{zinst}) is precisely the series expansion of the hypergeometric function ${}_2F_1(A,B;C;y)$, which solves the differential equation (\ref{hyper}) with $y=z$, this shows that the instanton partition function agrees with the chiral block (up to an overall factor). 

The matrix integral corresponding to a correlation function with $k$ insertions of $V_{-b/2}$  \textit{i.e.}
\be \label{LNpt}
\int \prod_{I} \D \la^I \prod_{I<J} |\la^I-\la^J|^{2\beta} \prod_{I}  (\la^I)^{ 2\al_1/\vep_1} (1-\la^I)^{2\al_2/\vep_1} \prod_{a=1}^{k} (z_a- \la^I)\,,
\ee
was also calculated exactly in \cite{Kaneko:1993}. The result is (proportional to) the generalised hypergeometric function\footnote{Note that it is also possible treat the cases with $\hat{V}_{-\ell b/2}$ insertions by setting $z_1=\ldots=z_\ell$ in (\ref{LNpt}) and   (\ref{genhyp}).}
\be \label{genhyp}
{}_{2}F^\bet_1(-N,\frac{1}{\bet}([2\al_1+2\al_2]/\vep_1+k+1)+N-1;\frac{1}{\bet}(2\al_1/\vep_1+k);z_1,\ldots,z_k)\,.
\ee
The function ${}_{2}F^\bet_1(A_1,A_2;B_1;z_1,\ldots,z_k)$ is defined as 
\be \label{hyperJack}
{}_{2}F^\bet_1(A,B;C;z_1,\ldots,z_k) 
= \sum_{\xi}  \frac{[A]^{\bet}_\xi[B]^{\bet}_\xi}{[C]^{\bet}_\xi} \frac{P^{\bet}_\xi(z_1,\ldots,z_k)}{|\xi|!}\,,
\ee
where the sum is over all partitions $\xi$ with at most $k$ parts,
\be
[X]_\xi^\bet = \prod_i (X-\frac{1}{\bet}(i-1))_{\xi_i}\,,
\ee
and $P_\xi^{\beta}(z_1,\ldots,z_k)$ is a (properly normalised) Jack polynomial. See \cite{Kaneko:1993} for further details about the notation. It should be possible to show that the corresponding CFT and gauge theory calculations lead to the same result.

We should also point out that the treatment of gauge theory surface and line operators in the Liouville language discussed in \cite{Alday:2009b} involves the insertion of vertex operators with $\al = -b/2$. This is precisely the class of operator insertions we have discussed above.

To go beyond the restricted set of correlation functions discussed above one possible approach would be to try to mimic the CFT method and determine the correlation functions in a perturbative expansion in the $z_i$. 

An alternative approach is to use matrix model perturbation theory. For the four--point function one can (at least when $\bet=1$) use the method in \cite{Dijkgraaf:2002d,Klemm:2002}; for the multi--cut and $A_r$ quiver extensions, see \textit{e.g.}~\cite{Naculich:2002}. One can also obtain a perturbative expansion within the framework of sections \ref{A1curve1} and \ref{A1curve2}; see section \ref{spert} below for some sample calculations. On the CFT side the matrix model perturbation theory is a somewhat peculiar expansion and it is not clear what its relevance is. A drawback of the perturbative matrix model approach is that solving for the stationary points lead to complicated expressions, but a clear advantage is that one can handle arbitrary punctures (\textit{i.e.}~arbitrary vertex operator momenta) using this method (both for the $A_1$ and $A_r$ theories). Let us also mention that in~\cite{Naculich:2002} another, less direct, matrix model approach to $\cN=2$ gauge theories was discussed; it might be interesting to try to connect it to the present approach.

\subsection{The curve: one--cut solutions} \label{A1curve1}

In this section we  analyse the one cut matrix model spectral curves, focusing on the matrix model corresponding to the three--point function discussed in section \ref{sA13pt}. We start by reviewing some well--known results. See \textit{e.g.} \cite{Marino:2004} for further details. In the standard diagonal gauge (eigenvalue basis) the one--matrix model partition function is\footnote{To conform with standard matrix model conventions $W$ in this section and in sections \ref{A1curve2} and \ref{spert} corresponds to $-W$ elsewhere in the paper. This also implies that the $\alpha_i$ differ by a sign.}
\bea
Z &=& \frac{1}{\left( 2\pi \right)^N N!} \int \prod_{I=1}^N \rmd \lambda^I  \prod_{I<J}(\la^I-\la^J)^2\, \exp \left(  -\frac{1}{g_s} \sum_{I=1}^N W(\lambda^I) \right) \nonumber \\
&=& \frac{1}{\left( 2\pi \right)^N N!} \int \prod_{I=1}^N \rmd \lambda^I\, \exp \left( N^2 S_{\mathrm{eff}} \left[ \rho(\lambda) \right] \right),
\eea
\noindent
where
\be
S_{\mathrm{eff}} \left[ \rho(\lambda) \right] =  -\frac{1}{t} \int_{\cC} \rmd\lambda\, \rho(\lambda) W(\lambda) + \iint_{\cC\times\cC} \rmd\lambda \rmd\lambda'\, \rho(\lambda) \rho(\lambda') \log \left|\lambda-\lambda'\right|\,.
\ee
Here $t$ is the 't~Hooft coupling $t=Ng_s$ and  we have introduced the eigenvalue density $\rho(\lambda) = \frac{1}{N} \sum_{I=1}^N \delta(\lambda-\lambda^I)$, normalized as $\int_{\cC} \rmd\lambda\, \rho(\lambda) = 1$. In this expression one still needs to specify the geometrical nature of the cut, $\cC$. In the most general case $\rho(\lambda)$ has compact support, with $\cC$ a multi--cut region with $s$ cuts. For the moment we shall focus on the one cut case, with $\cC = [a,b]$. If one now considers the Riemann surface which corresponds to a double--sheet covering of the complex plane, $\BC$, with precisely the above cut, it is  natural to define the $A$--cycle as the cycle around the cut. In this case, the $B$--cycle goes from the endpoint of the cut to infinity on one of the two sheets and back again on the other.

The generator of single--trace correlation functions is given by the resolvent
\be\label{resolvent}
\omega (z) = \frac{1}{N} \left\langle \tr\, \frac{1}{z-\Phi} \right\rangle = \frac{1}{N} \sum_{k=0}^{+\infty} \frac{1}{z^{k+1}} \left\langle \tr\, \Phi^k \right\rangle,
\ee
\noindent
which has the standard expansion $\omega(z) = \sum_{g=0}^{+\infty} g_s^{2g}\, \omega_g (z)$ with
\be
\omega_0 (z) = \int \rmd\lambda\, \frac{\rho(\lambda)}{z-\lambda}.
\ee
\noindent
The normalization of the eigenvalue density then implies that $\omega_0(z) \sim \frac{1}{z}$ as $z \to + \infty$. Also, observe that $\omega_0(z)$ is singular for $z \in \cC$ while it is analytic for $z \not\in \cC$. One may compute $\om_0(z)$ by making use of the large $N$ saddle--point equations of motion of the matrix model,
\be
\omega_0 (z+\rmi\epsilon) + \omega_0 (z-\rmi\epsilon) = \frac{1}{t} W'(z) = 2\, \mathsf{PV} \int_{\cC} \rmd\lambda\, \frac{\rho(\lambda)}{z-\lambda}.
\ee
\noindent
In a similar fashion, $\om_0(z)$ is  related to the eigenvalue density as
\be
\rho(z) = - \frac{1}{2\pi\rmi} \left( \omega_0 (z+\rmi\epsilon) - \omega_0 (z-\rmi\epsilon) \right) = - \frac{1}{\pi}\, \im\, \omega_0 (z).
\ee
\noindent
For a generic one--cut solution, the large $N$ resolvent is given by the ansatz
\be\label{genus0resolvent}
\omega_0(z) = \frac{1}{2t} \oint_{\cC} \frac{\rmd w}{2\pi\rmi}\, \frac{W'(w)}{z-w}\, \sqrt{\frac{(z-a)(z-b)}{(w-a)(w-b)}},
\ee
\noindent
where one still needs to specify the endpoints of the cut, $\{a,b\}$. An equivalent way to describe the matrix model geometry is via the corresponding spectral curve, $y(z)$, which basically describes the geometry of the Riemann surface we mentioned above. One may write
\be \label{M}
y(z) = W'(z) - 2t\, \omega_0(z) \equiv M(z)\, \sqrt{(z-a)(z-b)},
\ee
\noindent
with\footnote{This particular expression only holds for polynomial potentials.}
\be
M(z) = \oint_{(0)} \frac{\rmd w}{2\pi\rmi}\, \frac{W'(1/w)}{1-wz}\, \frac{1}{\sqrt{(1 - a w)(1 - b w)}},
\ee 
\noindent
where, again, one needs to specify the endpoints of the cut, $\{a,b\}$. The aforementioned large $z$ asymptotics of the resolvent immediately yield $2$ conditions for these $2$ unknowns. They are
\be\label{asympcond}
\oint_{\cC} \frac{\rmd w}{2\pi\rmi}\, \frac{w^n W'(w)}{\sqrt{(w-a)(w-b)}} = 2t\, \delta_{ns},
\ee
\noindent
for $n=0,1$, fully determining the endpoints of the single cut $s=1$.

It is also useful to define the holomorphic effective potential as $V_{\mathrm{h;eff}}'(z) = y(z)$. The effective potential is then given by the real part of the holomorphic effective potential, in such a way that
\be
V_{\mathrm{eff}}(\lambda) = \re \int^\lambda_a \rmd z\, y(z).
\ee
\noindent
The real part of the spectral curve therefore corresponds to the force exerted on a given eigenvalue. The imaginary part of the spectral curve, on the other hand, is related to the eigenvalue density via
\be
\rho(z) = \frac{1}{2\pi t}\, \im\, y(z).
\ee
\noindent
Finally, it turns out that one may also write the 't~Hooft parameter in terms of the spectral geometry as
\be
t = \frac{1}{4\pi\rmi} \oint_{A} \rmd z\, y(z) \,.
\ee

We now turn to our main point and consider the large $N$ expansion of the matrix model with potential 
\be \label{Where}
W (z) = \sum_{i=1}^{k} 2\alpha_i \log \left( z - z_i \right)
\ee
where $k$, $\{\alpha_i\}_{i=1}^k$ and $\{z_i\}_{i=1}^k$ are parameters we shall keep unspecified for the moment. It follows from (\ref{Where}) that
\be \label{W'}
W' (z) = \sum_{i=1}^{k} \frac{2\alpha_i}{z-z_i},
\ee
\noindent
implying that the logarithmic terms in (\ref{Where}) will not be terribly problematic --- one only needs to take into account extra poles, when moving the contours of integration around the complex plane. We begin by focusing on the one--cut solution, for which the large $N$ resolvent is given by the ansatz
\be
\omega_0(z) = \frac{1}{2t} \oint_{\cC} \frac{\rmd w}{2\pi\rmi}\, \frac{W'(w)}{z-w}\, \sqrt{\frac{(z-a)(z-b)}{(w-a)(w-b)}},
\ee
\noindent
where the integrand now has poles at the locations $\{z_i\}$ but, because the potential is purely logarithmic, any pole of the integrand at infinity is gone. In this case, a straightforward deformation of the integration contour reduces the integral along the cut to a sum of simple poles as
\be
\omega_0(z) = \frac{1}{2t} \left( W'(z) - \sum_{i=1}^{k} \frac{2\alpha_i}{\left( z-z_i \right) \sqrt{(z_i-a)(z_i-b)}}\, \sqrt{(z-a)(z-b)} \right).
\ee
\noindent
The large $z$ asymptotics, $\omega_0 (z) \sim \frac{1}{z} + \cdots$ as $z \to \infty$,  immediately implies that the endpoints of the cut $\cC=[a,b]$ are determined by the system
\bea
\sum_{i=1}^{k} \frac{2\alpha_i}{\sqrt{(z_i-a)(z_i-b)}} &=& 0, \\
\sum_{i=1}^{k} \left( 2\alpha_i - \frac{2\alpha_i z_i}{\sqrt{(z_i-a)(z_i-b)}} \right) &=& 2t,
\eea
\noindent
and the single--cut spectral geometry is then described by the curve
\be \label{singley}
y(z) = \sum_{i=1}^{k} \frac{2\alpha_i}{\left( z-z_i \right) \sqrt{(z_i-a)(z_i-b)}}\, \sqrt{(z-a)(z-b)}.
\ee
\noindent
One may also compute the holomorphic effective potential in a simple manner. We obtain
\bea
V_{\mathrm{h;eff}} (z) &=& - 2 \left( 2t - \sum_{i=1}^{k} 2\alpha_i \right) \log \left[ 2 \left( \sqrt{z-a} + \sqrt{z-b} \right) \right] + \\
&+&
 \sum_{i=1}^{k} \left( 2\alpha_i\, \log \left[ 1 - \frac{\sqrt{z-a}\, \sqrt{z_i-b}}{\sqrt{z-b}\, \sqrt{z_i-a}} \right] - 2\alpha_i\, \log \left[ 1 + \frac{\sqrt{z-a}\, \sqrt{z_i-b}}{\sqrt{z-b}\, \sqrt{z_i-a}} \right] \right), \nonumber
\eea
\noindent
with $a$ and $b$ determined by the system above. The structure of Stokes lines for this effective potential will be more complicated than in the usual polynomial cases.

Let us now specialise to the matrix model corresponding to the three--point function, with potential
\be \label{W3}
W_{\mathrm{3pf}} (z) = 2\alpha_1 \log z + 2\alpha_2 \log \left( z-1 \right)\,,
\ee
\noindent
and the further constraint
\be\label{t3pf}
t = \alpha_0 + \sum_{i=1}^{k} \alpha_i.
\ee
\noindent
Next we turn to the study of the  spectral geometry associated with the matrix model potential (\ref{W3}) beginning with the ``classical'' geometry.   The critical points are located at the points $z_*$ such that
\be
W_{\mathrm{3pf}}' (z_*) = \frac{2\alpha_1}{z_*} + \frac{2\alpha_2}{z_*-1} = 0.
\ee
\noindent
In this case, the general solution is
\be
z_* = \frac{\alpha_1}{\alpha_1+\alpha_2}.
\ee
\noindent
In the classical limit where the 't~Hooft coupling vanishes (\textit{i.e.}, where we choose vertex operators such that $\alpha_0+\alpha_1+\alpha_2=0$), one simply has $y(z) = W_{\mathrm{3pf}}'(z)$ and the cut collapses to the critical point of the potential, $[a,b] \to z_*$. 
In this particular case, 
\be
W_{\mathrm{3pf}} (z_*) = 2\alpha_1 \log \frac{\alpha_1}{\alpha_1+\alpha_2} + 2\alpha_2 \log \frac{\alpha_2}{\alpha_1+\alpha_2}.
\ee
\noindent
Now, the full spectral geometry will be such that the critical point $z_*$ opens up into a branch cut, of size $t$ (and not touching the marked points associated with the vertex operator insertions at $\{0,1,\infty\}$). This blown--up geometry of the spectral curve will have its shape determined by the parameters $\alpha_0$, $\alpha_1$ and $\alpha_2$. Clearly, because there is a single critical point, the spectral geometry will correspondingly have a single cut---the situation we studied above. The spectral geometry associated to the matrix model with potential (\ref{W3}) is therefore a genus zero one--cut Riemann surface.
From our previous general results it follows that the spectral curve is
\be 
y(z) = \left( \frac{2\alpha_1}{z \sqrt{ab}} + \frac{2\alpha_2}{\left( z-1 \right) \sqrt{(1-a)(1-b)}} \right) \sqrt{(z-a)(z-b)},
\ee
\noindent
where the endpoints $a$ and $b$ are obtained from the solution to the system
\bea \label{3ptconds}
\frac{2\alpha_1}{\sqrt{ab}} + \frac{2\alpha_2}{\sqrt{(1-a)(1-b)}} &=& 0, \non \\
\frac{2\alpha_2}{\sqrt{(1-a)(1-b)}} &=& -2\alpha_0,
\eea
\noindent
with (partial) solution
\bea
\sqrt{ab} &=& \frac{\alpha_1}{\alpha_0}, \\
\sqrt{(1-a)(1-b)} &=& - \frac{\alpha_2}{\alpha_0}.
\eea
\noindent
This immediately simplifies the spectral curve to
\be \label{curve2}
y(z) = \frac{2\alpha_0}{z \left( 1-z \right)}\, \sqrt{(z-a)(z-b)},
\ee
For completeness we also give the explicit solution to (\ref{3ptconds}):
\bea \label{ab}
\!\!\!\!\!\!\!\!\!a\! &\!=& \!\!\!\frac{\left( \alpha_0^2 {+} \alpha_1^2 {-} \alpha_2^2 {-}  \sqrt{ (\al_0 {-} \al_1 {-} \al_2) (\al_0 {+} \al_1 {-} \al_2) (\al_0{-} \al_1 {+} \al_2) (\al_0 {+} \al_1 {+} \al_2) } \right) }{2 \alpha_0^2}, \non\\
\!\!\!\!\!\!\!\!\!\! b &\!=& \!\!\!\frac{\left( \alpha_0^2 {+} \alpha_1^2 {-} \alpha_2^2 {+} \sqrt{ (\al_0 {-} \al_1 {-} \al_2) (\al_0 {+} \al_1 {-} \al_2) (\al_0{-} \al_1 {+} \al_2) (\al_0 {+} \al_1 {+} \al_2) } \right) }{2 \alpha_0^2} .
\eea
\noindent The spectral curve (\ref{curve2}) can be written
\be
y^2 = \frac{P_2(z)}{z^2(1-z)^2}\,,
\ee
where $P_2(z)$ is a polynomial of degree $2$. This expression was also written in \cite{Dijkgraaf:2009}; here we have also explicitly determined the coefficients in $P_2(z)$ in terms of the three $\al_i$'s.

\subsection{The curve: multi--cut solutions} \label{A1curve2}

\noindent
Let us return to the geometrical nature of the cut, $\cC$, where $\cC$ is now a multi--cut region with $s$ cuts. There are two cases: when $s$ is smaller or equal to the number of minima of the potential $V(\lambda)$, a typical situation in the standard matrix model context; or when $s$ is equal to the number of non--degenerate extrema of the potential $V(\lambda)$, the situation which arises when dealing with topological strings which is the case relevant to us. More precisely
\be
\cC = \bigcup_{I=1}^s A^I,
\ee
\noindent
where $A^I = [ x_{2I-1}, x_{2I} ]$ are the $s$ cuts and $x_1 < x_2 < \cdots < x_{2s}$. If one now considers the hyperelliptic Riemann surface which corresponds to a double--sheet covering of the complex plane, $\BC$, with precisely the same cuts as above, $A^I$, it is then natural to define the $A^I$--cycle as the cycle around the $A^I$ cut, with the $B_I$--cycle following via $B_I \cap A^J = \delta_I^J$. In this case, the $B_I$--cycle goes from the endpoint of the $A^I$ cut to infinity on one of the two sheets and back again on the other.

For a generic multi--cut solution, the large $N$ resolvent is given by the ansatz
\be\label{multi-genus0resolvent}
\omega_0(z) = \frac{1}{2t} \oint_{\cC} \frac{\rmd w}{2\pi\rmi}\, \frac{W'(w)}{z-w}\, \sqrt{\prod_{k=1}^{2s} \frac{z-x_k}{w-x_k}},
\ee
\noindent
where one still needs to specify the endpoints of the $s$ cuts, $\{x_k\}$. An equivalent way to describe the matrix model geometry is via the corresponding spectral curve, $y(z)$, which basically describes the hyperelliptic geometry of the Riemann surface we mentioned above. One may write
\be
y(z) = W'(z) - 2t\, \omega_0(z) \equiv M(z)\, \sqrt{\varsigma_s (z)},
\ee
\noindent
where
\be
\varsigma_s (z) \equiv \prod_{k=1}^{2s} (z-x_k)
\ee
\noindent
and\footnote{This particular expression only holds for polynomial potentials.}
\be
M(z) = \oint_{(0)} \frac{\rmd w}{2\pi\rmi}\, \frac{W'(1/w)}{1-wz}\, \frac{w^{s-1}}{\sqrt{\prod_{k=1}^{2s} (1 - x_k w)}},
\ee
\noindent
and where, again, one still needs to specify the endpoints of the $s$ cuts, $\{x_k\}$. The aforementioned large $z$ asymptotics of the resolvent immediately yield $s+1$ conditions for these $2s$ unknowns. They are
\be\label{multi-asympcond}
\oint_{\cC} \frac{\rmd w}{2\pi\rmi}\, \frac{w^n W'(w)}{\sqrt{\prod_{k=1}^{2s} (w-x_k)}} = 2t\, \delta_{ns},
\ee
\noindent
for $n=0,1,\ldots,s$. In order to fully solve the problem, one still requires $s-1$ extra conditions\footnote{Observe that no further conditions were required in the previous one--cut case, where $\cC=[x_1,x_2] \equiv [a,b]$. Indeed, in that situation the large $z$ asymptotics fully determined the endpoints of the single cut.} for the full set of $\{x_k\}$. These extra conditions depend on whether one wants to consider the standard matrix model or the topological string case. In the first option one considers all the different cuts at equipotential lines, where this condition may be written as
\be
\int_{x_{2I}}^{x_{2I+1}} \rmd z\, y(z) = 0.
\ee
\noindent
A physical understanding of this expression says that there is no force moving eigenvalues from one cut to another. In contrast, the topological string option (which is the case relevant to our analysis) generically corresponds to an unstable situation from a purely matrix model point of view. In this case one considers the filling fractions, 
\be
\eta^I \equiv \frac{N_I}{N} \equiv \int_{A^I} \rmd\lambda\, \rho(\lambda), \qquad I = 1,2,\ldots,s,
\ee
\noindent
as parameters, or moduli, of the problem under consideration. Observe that here $\sum_{I=1}^s \eta^I = 1$, making it an actual total of $s-1$ extra parameters, precisely the number required. By re--writing the eigenvalue density in terms of the resolvent, and the resolvent in terms of the spectral curve, one is led to the equivalent definition
\be
\eta^I = \frac{1}{4\pi\rmi t} \oint_{A^I} \rmd z\, y(z).
\ee
\noindent
One may also use as moduli the partial 't~Hooft couplings $t^I = t \eta^I = g_s N_I$. In this case 
\be
t^I = \frac{1}{4\pi\rmi} \oint_{A^I} \rmd z\, y(z),
\ee
\noindent
with $\sum_{I=1}^s t^I = t$, making a total of $s-1$ moduli.

Let us now return to the large $N$ expansion of the matrix model with the potential (\ref{Where}) and associated derivative (\ref{W'}) and briefly discuss how one may use standard saddle--point techniques to address multi--cut solutions. Again the logarithmic terms are not a problem; as we have seen before one only needs to take into account extra poles, when moving around the complex plane. When addressing multi--cut solutions, with $s$ cuts, the large $N$ resolvent is given by the ansatz (\ref{multi-genus0resolvent}), where the integrand now has poles at the locations $z_i$ and any pole of the integrand at infinity is gone. As in the single--cut case, a straightforward deformation of the integration contour reduces the integral along the cut to a sum of simple poles as
\be
\omega_0(z) = \frac{1}{2t} \left( W'(z) - \sum_{i=1}^{k} \frac{2\alpha_i}{\left( z-z_i \right) \sqrt{\prod_{n=1}^{2s} \left( z_i-x_n \right)}}\, \sqrt{\prod_{n=1}^{2s} \left( z-x_n \right)} \right).
\ee
\noindent
The large $z$ asymptotics, $\omega_0 (z) \sim \frac{1}{z} + \cdots$ as $z \to \infty$, immediately yield $s+1$ conditions on the $2s$ endpoints of the cuts via the system
\be
\sum_{i=1}^k \oint_{\cC} \frac{\rmd w}{2\pi\rmi}\, \frac{2\alpha_i\, w^m}{\left( w-z_i \right) \sqrt{\prod_{n=1}^{2s} (w-x_n)}} = 2t\, \delta_{ms},
\ee
\noindent
for $m=0,1,\ldots,s$. This may be written more explicitly as (notice that now the integrand does have a pole at infinity, when $m=s$)
\bea
\sum_{i=1}^{k} \frac{2\alpha_i}{\sqrt{\prod_{n=1}^{2s} \left( z_i-x_n \right)}} &=& 0, \non \\
\sum_{i=1}^{k} \frac{2\alpha_i\, z_i}{\sqrt{\prod_{n=1}^{2s} \left( z_i-x_n \right)}} &=& 0, \non \\[8pt]
&\cdots& \nonumber \\[8pt]
\sum_{i=1}^{k} \frac{2\alpha_i\, z_i^{s-1}}{\sqrt{\prod_{n=1}^{2s} \left( z_i-x_n \right)}} &=& 0, \\
\sum_{i=1}^{k} \left( 2\alpha_i - \frac{2\alpha_i\, z_i^s}{\sqrt{\prod_{n=1}^{2s} \left( z_i-x_n \right)}} \right) &=& 2t. \non 
\eea
\noindent
Of course in order to fully solve the problem one still requires $s-1$ extra conditions for the full set of endpoints $\{x_k\}$. Finally, the spectral geometry is described by the hyperelliptic spectral curve
\be
y(z) = \sum_{i=1}^{k} \frac{2\alpha_i}{\left( z-z_i \right) \sqrt{\prod_{n=1}^{2s} \left( z_i-x_n \right)}}\, \sqrt{\prod_{n=1}^{2s} \left( z-x_n \right)}.
\ee

Having understood the multi--cut spectral geometry we now focus on the case corresponding to the chiral four--point function in the Liouville theory, namely the matrix model with potential\footnote{Here and in section \ref{spert} for clarity we use $\zeta\equiv z_1$ to denote the location of the vertex operator insertion; in other sections $z$ is used.}
\be \label{W4}
W_{\mathrm{4pf}} (z) = 2\alpha_1 \log z + 2\alpha_2 \log \left( z-1 \right) + 2\alpha_3 \log \left( z-\zeta \right).
\ee
We begin with the ``classical'' geometry of the potential. The critical points are located at the points $z_*$ such that
\be
W_{\mathrm{4pf}}' (z_*) = \frac{2\alpha_1}{z_*} + \frac{2\alpha_2}{z_*-1} + \frac{2\alpha_3}{z_*-\zeta}= 0.
\ee
\noindent
In this case, the general solutions are
\bea
\!\!\!\!\!z_{*,1} &\!\!=&\!\! \frac{\left( 1{+}\zeta \right) \alpha_1 {+} \zeta \, \alpha_2 {+} \alpha_3 {-} \sqrt{\left( \left( 1{+}\zeta \right) \alpha_1 {+} \zeta\, \alpha_2 {+} \alpha_3 \right)^2 {-} 4 \alpha_1\, \zeta \left( \alpha_1{+}\alpha_2{+}\alpha_3 \right)}}{2 \left( \alpha_1+\alpha_2+\alpha_3 \right)}, \non \\
\!\!\!\!\!z_{*,2} &\!\!=&\!\! \frac{\left( 1{+}\zeta \right) \alpha_1 {+} \zeta\, \alpha_2 {+} \alpha_3 {+} \sqrt{\left( \left( 1+\zeta \right) \alpha_1 {+} \zeta\, \alpha_2 {+} \alpha_3 \right)^2 {-} 4 \alpha_1 \, \zeta \left( \alpha_1{+}\alpha_2{+}\alpha_3 \right)}}{2 \left( \alpha_1{+}\alpha_2{+}\alpha_3 \right)}.
\eea
\noindent
Generically, the critical points $z_{*,1}$ and $z_{*,2}$ will open up into branch cuts, of sizes $t_1$ and $t_2$ (and not touching the marked points associated to the vertex operator insertions at $\{0,1,\zeta,\infty\}$). Because of the two critical points, the most general spectral geometry will correspondingly have two--cuts and the spectral geometry associated with the chiral Liouville four--point function is a genus one two--cut (elliptic) Riemann surface. This two--cut blown--up geometry of the spectral curve will have its shape determined by $\alpha_0$, $\alpha_1$, $\alpha_2$ and $\alpha_3$. To be more precise, from the large $z$ asymptotics of the genus zero resolvent one obtains 3 conditions on the endpoints of the two cuts; the remaining required condition arising from the partial 't~Hooft moduli, $t_1$ or $t_2$ (where $t_1+t_2=t$). In \cite{Dijkgraaf:2009} this modulus is actually traded for $a=t_2-t_1$, the Coulomb modulus in the gauge theory, and we shall use this notation henceforth.

Of course there are particular points in the moduli space of the elliptic spectral curve where the geometry simplifies. One is the degenerate case where both partial 't~Hooft couplings vanish. Another special point occurs when only one of the critical points opens up into a branch cut, in which case one is dealing with a one--cut pinched spectral geometry, the pinch at the location of the critical point that remains ``closed''.  Let us consider this special case, where  
\bea \label{t1t2}
t_1 &=& - \frac{1}{2}\, a + \frac{1}{2} \sum_{i=0}^3 \alpha_i = 0, \\
t_2 &=& \frac{1}{2}\, a + \frac{1}{2} \sum_{i=0}^3 \alpha_i = t.
\eea
\noindent
From our previous (single--cut) result (\ref{singley}) we have
\bea
y(z) &=& \sqrt{(z-a)(z-b)} 
\non \\
&\times& \!\!\left[ \frac{2\alpha_1}{z \sqrt{ab}} + \frac{2\alpha_2}{\left( z-1 \right) \sqrt{(1-a)(1-b)}} + \frac{2\alpha_3}{\left( z-\zeta \right) \sqrt{(\zeta-a)(\zeta-b)}} \right] \!,
\eea
\noindent
where the endpoints $a$ and $b$ are a solution to the system
\bea
\frac{2\alpha_1}{\sqrt{ab}} + \frac{2\alpha_2}{\sqrt{(1-a)(1-b)}} + \frac{2\alpha_3}{\sqrt{(\zeta-a)(\zeta-b)}} &=& 0, \\
\frac{2\alpha_2}{\sqrt{(1-a)(1-b)}} + \frac{\zeta\, 2\alpha_3}{\sqrt{(\zeta-a)(\zeta-b)}} &=& - 2\alpha_0.
\eea
\noindent
This may be equivalently written as
\bea
\frac{2\alpha_1}{\sqrt{ab}} &=& 2\alpha_0 + \frac{\left( \zeta-1 \right) 2\alpha_3}{\sqrt{(\zeta-a)(\zeta-b)}}, \\
\frac{2\alpha_2}{\sqrt{(1-a)(1-b)}} &=& - 2\alpha_0 - \frac{\zeta 2\alpha_3}{\sqrt{(\zeta-a)(\zeta-b)}},
\eea
\noindent
which slightly simplifies the spectral curve to
\be
y(z) = \left( \frac{2\alpha_0}{z \left( 1-z \right)} + \frac{\zeta \left( \zeta-1 \right)}{z \left( z-1 \right) \left( z-\zeta \right)}\, \frac{2\alpha_3}{\sqrt{(\zeta-a)(\zeta-b)}} \right) \sqrt{(z-a)(z-b)}.
\ee
\noindent
Generically, however, we are dealing with a system of quartic equations and, although it can be solved algebraically, its exact solution is not terribly illuminating. In the following, we therefore choose a different route and solve this system perturbatively in $\zeta$. This is motivated by the expansion on the CFT side and simplifies the problem considerably. To first order, we obtain the solution
\bea
\sqrt{a b} &\!=& \!\frac{\alpha_1+\alpha_3}{\alpha_0} - \frac{\left( \left( \alpha_1+\alpha_3 \right)^2 + \alpha_2^2 - \alpha_0^2 \right) \alpha_3\, \zeta}{2 \alpha_0 \left( \alpha_1+\alpha_3 \right)^2} + \cO \left( \zeta^{3/2} \right), \non \\
\sqrt{\left( 1-a \right) \left( 1-b \right)} &\!=&\! - \frac{\alpha_2}{\alpha_0} \left( 1 - \frac{\alpha_3 \,\zeta}{\alpha_1+\alpha_3} + \cO \left( \zeta^2 \right) \right),
\eea
\noindent
or, equivalently,
\bea
a &=& \frac{\alpha_0^2 {+} \left( \alpha_1{+}\alpha_3 \right)^2 {-} \alpha_2^2 {-} \sqrt{\Omega}}{2 \alpha_0^2} + \frac{\alpha_0^2 {-} \left( \alpha_1{+}\alpha_3 \right)^2 {+} \alpha_2^2 {+} \sqrt{\Omega}}{2 \alpha_0^2 \left( \alpha_1+\alpha_3 \right)}\, \alpha_3 \zeta + \cO \left( \zeta^2 \right), \non \\
b &=& \frac{\alpha_0^2 {+} \left( \alpha_1{+}\alpha_3 \right)^2 {-} \alpha_2^2 {+} \sqrt{\Omega}}{2 \alpha_0^2} {+} \frac{\alpha_0^2 {-} \left( \alpha_1{+}\alpha_3 \right)^2 + \alpha_2^2 {-} \sqrt{\Omega}}{2 \alpha_0^2 \left( \alpha_1+\alpha_3 \right)}\, \alpha_3 \zeta + \cO \left( \zeta^2 \right),
\eea
\noindent
where we have defined
\be\label{Omega}
\Omega \equiv \left( \alpha_0{+}\alpha_1{+}\alpha_2{+}\alpha_3 \right) \left( \alpha_0{-}\alpha_1{+}\alpha_2{-}\alpha_3 \right) \left( \alpha_0{+}\alpha_1{-}\alpha_2{+}\alpha_3 \right) \left( \alpha_0{-}\alpha_1{-}\alpha_2{-}\alpha_3 \right).
\ee
This solution will be important in the perturbative calculations in section \ref{spert} below.

From having worked out some degenerate cases of the matrix model associated with the Liouville four--point function we have acquired some intuition about what to expect as we move on to the case with arbitrary vertex operators with conformal dimensions such that, generically, one will find a two--cut geometry. Let us briefly comment on this geometry. We have previously computed the spectral curve for a general multi--cut case and, for the two--cut ansatz associated to the Liouville four--point function, this is
\bea
y(z) &=& \left( \frac{2\alpha_1}{z \sqrt{x_1 x_2 x_3 x_4}} + \frac{2\alpha_2}{\left( z-1 \right) \sqrt{\left( 1-x_1 \right) \left( 1-x_2 \right) \left( 1-x_3 \right) \left( 1-x_4 \right)}} + \right. \nonumber \\
&&
\left. + \frac{2\alpha_3}{\left( z-\zeta \right) \sqrt{\left( \zeta-x_1 \right) \left( \zeta-x_2 \right) \left( \zeta-x_3 \right) \left( \zeta-x_4 \right)}} \right) \sqrt{\varsigma(z)}
\eea
\noindent
with
\be
\varsigma(z) = \left( z-x_1 \right) \left( z-x_2 \right) \left( z-x_3 \right) \left( z-x_4 \right).
\ee
\noindent
It is simple to see that the endpoints of the two cuts, $\{x_i\}_{i=1}^4$, are now a solution to the system
\bea
\sqrt{x_1 x_2 x_3 x_4} &=& - \frac{\zeta\, \alpha_1}{\alpha_0}, \\
\sqrt{\left( 1-x_1 \right) \left( 1-x_2 \right) \left( 1-x_3 \right) \left( 1-x_4 \right)} &=& \frac{\left( \zeta-1 \right) \alpha_2}{\alpha_0}, \\
\sqrt{\left( \zeta-x_1 \right) \left( \zeta-x_2 \right) \left( \zeta-x_3 \right) \left( \zeta-x_4 \right)} &=& - \frac{\left( \zeta-1 \right) \zeta\, \alpha_3}{\alpha_0},
\eea
\noindent
which immediately simplifies the two--cut spectral curve as
\be
y(z) = - \frac{2\alpha_0}{z \left( z-1 \right) \left( z-\zeta \right)}\, \sqrt{\varsigma(z)}.
\ee
\noindent
Further notice, as we have discussed before, that above we have 3 equations for 4 unknowns, and there is still one further moduli to consider; either $t_1$ or $t_2$ (they are not independent as $t_1+t_2=t$),
\bea
t_1 &=& - \frac{1}{2}\, a + \frac{1}{2} \sum_{i=0}^3 \alpha_i, \\
t_2 &=& \frac{1}{2}\, a + \frac{1}{2} \sum_{i=0}^3 \alpha_i.
\eea

\subsection{Some perturbative calculations} \label{spert}

We shall now discuss some perturbative calculations in the matrix models considered above. By perturbative, we mean a 't Hooft expansion in $g_s$. As a warm up we focus on the matrix model (\ref{W3}). One can compute the partition function exactly (see appendix \ref{ortho}); the result can be written 
\be
Z = \frac{G_2 \left( N+1 \right)}{\left( 2\pi \right)^N} \frac{G_2 \left( -N-2\alpha_0/g_s+1 \right)}{G_2 \left( -2\alpha_0/g_s+1 \right)} \frac{G_2 \left( N+2\alpha_1/g_s+1 \right)}{G_2 \left( 2\alpha_1/g_s+1 \right)}\, \frac{G_2 \left( N+2\alpha_2/g_s+1 \right)}{G_2 \left( 2\alpha_2/g_s+1 \right)}.
\ee
\noindent where $G_2(z)$  is the Barnes function. The Barnes function has the asymptotic expansion 
\bea
\log G_2 (N+1) &=& \frac{1}{2} N^2 \log N + \frac{1}{2} N \log 2\pi - \frac{3}{4} N^2 - \frac{1}{12} \log N + \zeta'(-1) + \nonumber \\
&& + \sum_{g=2}^{+\infty} \frac{1}{N^{2g-2}}\, \frac{B_{2g}}{2g(2g-2)},
\label{barnesexpansion}
\eea
\noindent
with $B_{2g}$ the Bernoulli numbers. However, (\ref{barnesexpansion}) only deals with the $N \to + \infty$, or $g_s \to 0^+$, asymptotic region. Depending on the sign of the finite parameters $\alpha_1$ and $\alpha_2$, as $g_s \to 0^+$ one will have $- \frac{\alpha_i}{g_s}$ either going to $+\infty$ or to $-\infty$, and one thus needs to also understand the asymptotics of the logarithm of the Barnes function in the region $N \to - \infty$. However, it turns out that, from the point of view of the perturbative expansion, this sign difference is not very relevant. To clarify this issue, first notice the relation
\be
\log G_2 \left( 1-N \right) = \log G_2 \left( 1+N \right) - N \log 2\pi + \int_0^N \rmd x\, \pi x \cot \pi x.
\ee
\noindent
Explicitly evaluating the integral we find
\bea
\log G_2 \left( 1-N \right) &=& \log G_2 \left( 1+N \right) - N \log 2\pi + \frac{\rmi \pi}{12} \left( 1-6N^2 \right) + \nonumber\\
&&
+ N \log \left( 1 - \rme^{2 \pi \rmi N} \right) - \frac{\rmi}{2\pi}\, {\mathrm{Li}}_2 \left( \rme^{2 \pi \rmi N} \right),
\eea
\noindent
where ${\mathrm{Li}}_2 (z)$ is the dilogarithm. The logarithmic and dilogarithmic contributions can be expanded as
\be
N\, \log \left( 1 - \rme^{2 \pi \rmi N} \right) - \frac{\rmi}{2\pi}\, {\mathrm{Li}}_2 \left( \rme^{2 \pi \rmi N} \right) = - \sum_{m=1}^{+\infty} \left( \frac{N}{m} - \frac{1}{2\pi \rmi m^2}\right) \rme^{2\pi\rmi N m}.
\ee
\noindent
As explained in \cite{Pasquetti:2009}, this is actually the instanton contribution to the Barnes function, also describable in terms of Stokes phenomena (across the $+ \frac{\pi}{2}$ Stokes line). In other words, this contribution is purely nonperturbative and we shall neglect it at this stage, \textit{i.e.}, from a \textit{purely perturbative} point of view we may  use following result
\be
\log G_2 \left( 1-N \right) \simeq \log G_2 \left( 1+N \right) - N \log 2\pi + \frac{\rmi \pi}{12} \left( 1-6N^2 \right).
\ee
\noindent

From the above discussion it follows that the logarithm of $Z$, in the 't Hooft limit, has the expansion\footnote{As usual, we are only considering the real part of the free energy.}
\be
F\equiv \log Z=\sum_{g} g_s^{2g-2} F_g\,,
\ee
where, if we define $\cF=F-\frac{1}{2}N^2\log g_s-\frac{1}{2}N\log2\pi+\log G_2(N+1)$ (essentially amounting to the Gaussian normalization of the free energy) and use the relation (\ref{t3pf}) with $k=2$, we find 
\bea
\cF^{\mathrm{3pf}}_0 &=& \frac{1}{2} \left( -\alpha_0+\alpha_1+\alpha_2 \right)^2 \left( \log [-\alpha_0+\alpha_1+\alpha_2] - \frac{3}{2} \right) - 2\alpha_0^2 \left( \log [2\alpha_0] - \frac{3}{2} \right)  \nonumber \\
&+&
 \frac{1}{2} \left( \alpha_0-\alpha_1+\alpha_2 \right)^2 \left( \log [\alpha_0-\alpha_1+\alpha_2] - \frac{3}{2} \right) - 2\alpha_1^2 \left( \log [2\alpha_1] - \frac{3}{2} \right)  \nonumber \\
&+&
 \frac{1}{2} \left( \alpha_0+\alpha_1-\alpha_2 \right)^2 \left( \log [\alpha_0+\alpha_1-\alpha_2] - \frac{3}{2} \right) - 2\alpha_2^2 \left( \log [2\alpha_2] - \frac{3}{2} \right)
\eea
\noindent
for $g=0$;
\be
\cF^{\mathrm{3pf}}_1 =  - \frac{1}{12} \log \frac{-\alpha_0+\alpha_1+\alpha_2}{2\alpha_0} - \frac{1}{12} \log \frac{\alpha_0-\alpha_1+\alpha_2}{2\alpha_1} - \frac{1}{12} \log \frac{\alpha_0+\alpha_1-\alpha_2}{2\alpha_2}
\ee
\noindent
for $g=1$; and
\bea
\cF^{\mathrm{3pf}}_g &=& \frac{ B_{2g}}{2g \left( 2g-2 \right)} \bigg[ \left( -\alpha_0+\alpha_1+\alpha_2 \right)^{2-2g} - \left( 2\alpha_0 \right)^{2-2g}  \\
&+&
 \left(\alpha_0-\alpha_1+\alpha_2\right)^{2-2g} - \left( 2\alpha_1 \right)^{2-2g} + \left(\alpha_0+\alpha_1-\alpha_2\right)^{2-2g} - \left( 2\alpha_2 \right)^{2-2g} \bigg] \nonumber
\eea
\noindent
for $g \ge 2$. As alluded to above, it is also rather straightforward to compute the full nonperturbative contribution to this result. Because this is far from our present discussion we refer the reader to \cite{Pasquetti:2009} for details, but the main idea essentially follows from the application of
\be
{\mathrm{disc}}\, \log G_2 \left( N+1 \right) = \rmi \sum_{m=1}^{+\infty} \left( \frac{|N|}{m} + \frac{1}{2\pi m^2} \right) \rme^{-2\pi |N| m}
\ee
\noindent
to the expression for the free energy (where the discontinuity of the free energy will yield the full tower of multi--instanton corrections). In this case one simply obtains
\bea
{\mathrm{disc}}\, \cF_{\mathrm{3pf}} &=& \frac{\rmi}{2\pi\bar{g}_s} \sum_{m=1}^{+\infty} \left( \frac{2\pi \left( -\alpha_0+\alpha_1+\alpha_2 \right)}{m} + \frac{\bar{g}_s}{m^2} \right) \rme^{-\frac{2\pi \left( -\alpha_0+\alpha_1+\alpha_2 \right) m}{\bar{g}_s}} + \nonumber \\
&+&
  \frac{\rmi}{2\pi\bar{g}_s} \sum_{m=1}^{+\infty} \left( \frac{2\pi \left( \alpha_0-\alpha_1+\alpha_2 \right)}{m} + \frac{\bar{g}_s}{m^2} \right) \rme^{-\frac{2\pi \left(\alpha_0-\alpha_1+\alpha_2\right) m}{\bar{g}_s}} + \nonumber \\
&+&
 \frac{\rmi}{2\pi\bar{g}_s} \sum_{m=1}^{+\infty} \left( \frac{2\pi \left( \alpha_0+\alpha_1-\alpha_2 \right)}{m} + \frac{\bar{g}_s}{m^2} \right) \rme^{-\frac{2\pi \left(\alpha_0+\alpha_1-\alpha_2\right) m}{\bar{g}_s}} - \nonumber \\
&-&
 \frac{\rmi}{2\pi\bar{g}_s} \sum_{m=1}^{+\infty} \left( \frac{4\pi\alpha_0}{m} {+} \frac{\bar{g}_s}{m^2} \right) \rme^{-\frac{4 \pi\alpha_0 m}{\bar{g}_s}} - \frac{\rmi}{2\pi \bar{g}_s} \sum_{m=1}^{+\infty} \left( \frac{4\pi\alpha_1}{m} {+} \frac{\bar{g}_s}{m^2} \right) \rme^{-\frac{4\pi \alpha_1 m}{\bar{g}_s}} - \nonumber \\
&-&
 \frac{\rmi}{2\pi \bar{g}_s} \sum_{m=1}^{+\infty} \left( \frac{4\pi\alpha_2}{m} {+} \frac{\bar{g}_s}{m^2} \right) \rme^{-\frac{4\pi \alpha_2 m}{\bar{g}_s}}.
\eea
\noindent
Notice that this is an exact result, to all loops and including all instanton numbers.

Even though we have (for this case) an exact perturbative expression it is useful to also compute $F_1$ using a method that generalises to more complicated situations where exact results are unavailable. There is a universal formula for $F_1$ which takes the form \cite{ACKM:1993}\footnote{The universal expression was derived for polynomial potentials, but our results indicate that it also holds for the multi--Penner--type potentials considered in \cite{Dijkgraaf:2009}.}
\be \label{univ}
F_1 = - \frac{1}{24} \log \left( M(a) M(b) (a-b)^4 \right)
\ee 
\noindent
where (this follows from (\ref{M}) and (\ref{ab}))
\be
M(a) M(b) = 4 \frac{\alpha_0^6}{\alpha_1^2 \alpha_2^2}.
\ee
\noindent
It immediately follows, using (\ref{ab}), that
\bea
F_1  &=& - \frac{1}{12} \log \left( 2\alpha_0+2\alpha_1+2\alpha_2 \right) - \frac{1}{12} \log \frac{-\alpha_0+\alpha_1+\alpha_2}{\alpha_0} - \nonumber \\
&&
- \frac{1}{12} \log \frac{\alpha_0-\alpha_1+\alpha_2}{\alpha_1} - \frac{1}{12} \log \frac{\alpha_0+\alpha_1-\alpha_2}{\alpha_2},
\label{F13pf}
\eea
\noindent
reproducing the result we have previously obtained, up to some irrelevant numerical terms (this expression explicitly includes the Gaussian contribution $-\frac{1}{12} \log t$).  

Next we turn to the case corresponding to the chiral four--point function in the Liouville theory, namely the matrix model with potential (\ref{W4}). In this case we do not have an exact solution, but as discussed in the section \ref{A1curve2} we can get tractable expressions if we work order by order in $\zeta$. 

As an example we consider the special case discussed at the end of section \ref{A1curve2} and focus on $F_1$, which may be computed in a straightforward fashion from the universal result (\ref{univ}). We have
\bea
M(z) &=& \frac{2\alpha_1}{z \sqrt{ab}} + \frac{2\alpha_2}{\left( z-1 \right) \sqrt{(1-a)(1-b)}} + \frac{2\alpha_3}{\left( z-\zeta \right) \sqrt{(\zeta-a)(\zeta-b)}} \non \\
&\simeq&
\frac{2\alpha_1+2\alpha_3}{z \sqrt{ab}} + \frac{2\alpha_2}{\left( z-1 \right) \sqrt{(1-a)(1-b)}} + \frac{2\alpha_3\, \zeta}{z^2 \sqrt{ab}} + \frac{2\alpha_3 \left( a+b \right) \zeta}{2 z \left( ab \right)^{\frac{3}{2}}} + \cO \left( \zeta^2 \right)  \nonumber \\
&=&
\frac{2\alpha_0}{z (1-z)} \left( 1 + \frac{\alpha_3 \,\zeta}{z \left( \alpha_1+\alpha_3 \right)} + \cO \left( \zeta^2 \right) \right), 
\eea
leading to
\be
M(a) M(b) = \frac{4\alpha_0^2}{a b (1-a) (1-b)} \left( 1 + \frac{a+b}{ab}\, \frac{\alpha_3\, \zeta}{\alpha_1+\alpha_3} + \cO \left( \zeta^2 \right) \right),
\ee
\noindent
and further computing
\be
b-a = \frac{\sqrt{\Omega}}{\alpha_0^2} \left( 1 - \frac{\alpha_3 \,\zeta}{\alpha_1+\alpha_3} + \cO \left( \zeta^2 \right) \right),
\ee
\noindent
where $\Omega$ was defined in (\ref{Omega}), and
\be
b+a = \frac{\alpha_0^2 + \left( \alpha_1+\alpha_3 \right)^2 - \alpha_2^2}{\alpha_0^2} + \frac{\alpha_0^2 - \left( \alpha_1+\alpha_3 \right)^2 + \alpha_2^2}{\alpha_0^2 \left( \alpha_1+\alpha_3 \right)}\, \alpha_3\, \zeta + \cO \left( \zeta^2 \right)
\ee
\noindent
it finally follows, after putting all the above expressions together,
\bea
M(a) M(b) &=& \frac{4\alpha_0^6}{\alpha_2^2 \left( \alpha_1+\alpha_3 \right)^2} \left( 1 + \frac{4\alpha_3\, \zeta}{\alpha_1+\alpha_3} + \cO \left( \zeta^2 \right) \right), \\
\left( b-a \right)^4 &=& \frac{\Omega^2}{\alpha_0^8} \left( 1 - \frac{4 \alpha_3\, \zeta}{2\alpha_1+2\alpha_3} + \cO \left( \zeta^2 \right) \right), 
\eea
and (here we also included the $\zeta^2$ terms)
\bea \label{mmF1}
&&\!\!\!\!\!\!\!\!F_1  = 
- \frac{1}{12} \log \frac{2\Omega}{\alpha_0 \left( \alpha_1+\alpha_3 \right) \alpha_2} 
 \\
&&\!\!\!\!\!\!\!\!\!\! -\frac{\al_1\al_3({-}\al_0{+}\al_1{-}\al_2{+}\al_3) (\al_0{+}\al_1{-}\al_2{+}\al_3) ({-}\al_0{+}\al_1{+}\al_2{+}\al_3)(\al_0{+}\al_1{+}\al_2{+}\al_3)}{32(\al_1+\al_3)^6}\zeta^2\,.  \non 
\eea
\noindent
Notice that, up to numerical constants that we drop, and to first non--trivial order in $\zeta$, this result is exactly the same as the result in (\ref{F13pf}), except for the shift $\alpha_1 \to \alpha_1+\alpha_3$. Note also that the order $\cO (\zeta)$ contribution to $F_1$ vanishes. 

The above result can be compared to the corresponding result for the Nekrasov instanton partition function of the $\SU(2)$ theory with four fundamental hypermultiplets. In this case one easily obtains
\be \label{NF1}
F^{\rm inst}_1 = \frac{ - a^6 \si_1(m^2) + 2 a^4 \si_2(m^2) -3 a^2 \si_3(m^2)+4 \si_4(m^2) }{128 a^8} y^2 + \cO(y^3)
\ee
where $\si_k(m^2) = \sum_{i_1<\cdots<i_k} m_{i_1}^2 \cdots m_{i_k}^2$. Note that the $\cO(y)$ term vanishes in agreement with (\ref{mmF1}) identifying $y$ with $\zeta$. Furthermore, after implementing the relations,
\be
m_1=\al_1+\al_3 \,, \quad m_2=\al_1-\al_3 \,, \quad m_3=\al_0+\al_2 \,, \quad m_4=\al_0-\al_2
\ee
together with $a = m_1$ we see that also the second order terms in (\ref{mmF1}) and (\ref{NF1}) agree perfectly. (The definition of $a$ in the Nekrasov expression differs from the one in (\ref{t1t2}).) This result is consistent with the analysis in section \ref{A1high} and supports the approach in \cite{Dijkgraaf:2009}.

Computing $\cF_g$ for $g\ge 2$ would require heavier machinery, see \textit{e.g.} \cite{Eynard:2007}, and will not be attempted here. Similarly, an explicit example involving the full--fledged two--cut geometry would take us too far afield. The integrals of the periods associated to the 't~Hooft moduli are generically hard to evaluate exactly (although it is possible to do so in the present situation) and even harder to invert in order to find explicit solutions for the endpoints of the two cuts. A possible way out is to resort to perturbation theory, along the lines of \cite{Dijkgraaf:2002d,Klemm:2002}, but we shall leave this question for future work.

Let us close with a comment about the relation to the chiral four--point function in the Liouville theory to leading order in $\zeta$.  In order to reconstruct the full Liouville three--point function at this order pertubative calculations are not enough, one would also need to add the full set of nonperturbative corrections and in general also let $\beta\neq 1$, in order to obtain the desired result. It seems plausible that a matrix model perturbation theory in $\zeta$ exists (for general $\bet$) which is exact in $g_s$  order--by--order, but we leave this question for future work.

\setcounter{equation}{0} 
\section{The $A_r$ matrix models} \label{sAr}

In this section we perform several calculations in the $A_r$ quiver matrix models. The resulting expressions are compared to the corresponding expressions in the $A_r$ Toda theories and the $A_r$ quiver gauge theories.

\subsection{The three--point function} \label{Arthree}

The above analysis of the $A_1$ matrix model three--point function (see section \ref{sA13pt}) can be extended to the $A_r$ theory for any $r$. The relevant integral is
\be  \label{SelAr}
S_r(\al_1,\al_2,\bet) = \int \prod_{iI} \D \la^I_i \prod_{(i,I)<(j,J)} |\la^I_i-\la_j^J|^{\beta A_{ij}}\prod_{i,I} (\la^I_i)^{ \al^i_1/\vep_1} \prod_{i,I} (1-\la_i^I)^{\al_2^i/\vep_1} \,.
\ee
Here $\al_1^i = \lb \al_1,e_i \rb$ with $\al_1 = \al_1^i \La_i$ and similarly for $\al_2$ (see appendix \ref{Alie} for our Lie algebra conventions). The integral (\ref{SelAr})  can be explicitly evaluated provided one imposes the restriction $\al_1^i = \vka \de_{i r}$ (\textit{i.e.}~$\al_1=\vka \La_r$); the result is \cite{Warnaar:2007}
\bea \label{Arsel}
&&\prod_{1\le i\le j\le r} \!\!\!\!\prod_{I=1}^{N_i-N_{i-1}} \!\!\!\!\frac{\Ga([\al_2^i+\cdots\al_2^j]/\vep_1 +j-i+1 + (I-1+i-j)\bet)}{\Ga([\al_2^i+\cdots\al_2^j + \al_1^j]/\vep_1 +j-i+2 + (I-2+i-j+N_j-N_{j+1})\bet)} \non \\ 
&&\times \prod_{i=1}^{ r} \prod_{I=1}^{N_i} \frac{\Ga(\al_1^i/\vep_1 + 1 + (I-N_{i+1}-1)\bet)\Ga(I\bet)}{\Ga(\bet)} 
\eea
where $N_0=N_{r+1}=0$. The result (\ref{Arsel}) depends on a very particular choice of integration contour, which is quite subtle and will not be discussed here; see \cite{Warnaar:2007} for further details. 

Using various Lie algebra results (see appendix \ref{Alie} for further details) one can show that the relation 
\be
\al^i_1 + \al^i_2 -\al^i_0 -\vep_2 A_{ij} N_j=0\,,
\ee
where  $\al_1=\vka \La_r$ (\textit{i.e.}~$\al_1^i=\vka\de_{ir}$) implies that 
\be
-\vep_2(N_i - N_{i+1}) = - A_{ir}^{-1}\vka +A_{i+1,r}^{-1}\vka+ \lb  \al_2, h_{i+1} \rb  -  \lb  \al_0, h_{i+1} \rb \,.
\ee
(Note the special case, $\vep_2 N_1 = \frac{\vka}{r+1}+ \lb  \al_2, h_{1} \rb  -  \lb  \al_0, h_{1} \rb$.) Furthermore,
\be
\al_1^i  - A_{ir}^{-1}\vka +A_{i+1,r}^{-1}\vka = \frac{\vka}{r+1}\,.
\ee
Using these results together with 
\be
\al_2^i+\cdots+\al_2^j =  - \lb  \al_2, u_{j+1} - u_i \rb \,,\qquad
i -j-1= -\lb \rho, u_{i}-u_{j+1} \rb\,,
\ee
and (\ref{Gaid}), it follows that the expression (\ref{Arsel}) can be written (we suppress an unimportant prefactor and $\Ga_2(x)$ is short--hand for $\Ga_2(x|-\vep_1,-\vep_2)$)
\bea \label{M3}
&&\! \! \! \! \! \!\prod_{1\le i < j \le r+1} \! \! \! \! \! \! \! \!  \frac{ \Ga_2(  \lb  \al_2{+}\rho \vep, h_{j} \rb  {-}  \lb  \al_0{+}\rho \vep, h_{i} \rb {+}  \frac{\vka}{r{+}1})\Ga_2( {-}\vep {-}\lb  \al_2{+}\rho \vep, h_{i} \rb  {+} \lb  \al_0{+}\rho \vep, h_{j} \rb {-}  \frac{\vka}{r{+}1})  } { \Ga_2( - \lb  \al_2+\rho \vep, e_{ij} \rb )\Ga_2( -\vep - \lb  \al_0+\rho \vep, e_{ij} \rb ) }\non \\
 &&\! \! \! \!\times  \frac{\Ga_2( {-}\vep {-}\lb  \al_2{+}\rho \vep, h_{1} \rb  {+} \lb  \al_0{+}\rho \vep, h_{1} \rb {-}  \frac{\vka}{r{+}1})}{\Ga_2(-\vep-\vka)}\prod_{i=2}^{r+1} \Ga_2( \lb  \al_2{+}\rho \vep, h_{i} \rb  {-} \lb  \al_0{+}\rho \vep, h_{i} \rb {+}  \frac{\vka}{r{+}1}) \non \\
 &&\! \! \! \!\times \prod_{i=2}^{r} \frac{\Ga_2(N_i\vep_2 ) }{\Ga_2(-\vep-N_i\vep_2)} \,.
\eea
The first thing to note is that the factors on the last line is a phase. As in the $A_1$ case discussed in section \ref{sA13pt} the above expression can be compared to the three--point function in the $A_r$ Toda theory. After suitably rescaling the vertex operators  with multiplicative factors depending on their momenta, the  $A_r$ Toda theory  three--point function (\ref{t3pt}) can be written 
\bea \label{T3}
 &&\prod_{i,j=1}^{r+1} \big[\Ups\big(\frac{\vka}{r+1}+
    \lb\al_1-Q\rho,h_i\rb+\lb\al_2-Q\rho,h_j \rb\big)\big]^{-1} \non \\
&=&     \prod_{1\le i<j\le r+1} \big| \Ga_b\big(\frac{\vka}{r+1}+
    \lb\al_1-Q\rho,h_i\rb+\lb\al_2-Q\rho,h_j \rb\big) \big|^2\non  \\
    &\times&  \prod_{1\le i< j \le r+1} \big|  \Ga_b\big(Q-\frac{\vka}{r+1}-
    \lb\al_1-Q\rho,h_j\rb-\lb\al_2-Q\rho,h_i \rb\big)\big|^2 \non \\
  &\times&  \big| \Ga_b\big(Q-\frac{\vka}{r+1}-
    \lb\al_1-Q\rho,h_1\rb-\lb\al_2-Q\rho,h_1 \rb\big) \big|^2  \\ 
    &\times& \prod_{i=2}^{ r+1} \big| \Ga_b\big(\frac{\vka}{r+1}+
    \lb\al_1-Q\rho,h_i\rb+\lb\al_2-Q\rho,h_i \rb\big) \big|^2 ,\non
\eea
where $\Ga_b(x)$ is short--hand for $\Ga_2(x|b,1/b)$. By similarly rescaling of the matrix model vertex operators we are left with the factors in the numerators on the first two lines of (\ref{M3}). After using the identification (\ref{epb}) together with $Q\rho-\al_0 \rar -(Q\rho-\al_1)$ we find complete agreement with the``square-root" of (\ref{T3}). The expression (\ref{M3}) should probably also be related to (the perturbative piece in) the $T_{r+1}=\cT_{3,0}(A_r)$ theory where one of the masses satisfies a restriction inherited from  $\al_1=\vka \La_r$ via the AGT relation. 

We note that  in \cite{Fateev:2005, Fateev:2007b} a complex version of the above integral was used to derive the three--point function in the $A_r$ Toda theory when one of the momenta takes the special value $\vka \La_r$.

In the above evaluation of the integral (\ref{SelAr}) the condition $\al_1=\vka \La_r$ was imposed with $\al_2$ left arbitrary. In  \cite{Warnaar:2009}  the above integral was evaluated for the rank $2$ case with the alternative restriction: $\al_2 = \vka \La_1 -(\vka+\vep)\La_2$, with $\al_1$ left arbitrary.

To understand why there is more than one possible choice which allows for an explicit evaluation of the above integral, we recall that in the $A_2$ Toda theory the condition $\al = \vka \La_r$ translates into the fact that the corresponding $\cW$ primary state satisfies (see \textit{e.g.}~\cite{Fateev:2007b,Wyllard:2009} and references therein)
\be
 \left(W_{-1}-\frac{3w(\al)}{2\De(\al)}L_{-1}\right)|\al\rb=0\,,
\ee
 which implies
\be \label{Dew}
\De(\al)^2\left(\frac{32}{22+5c}\left[\De(\al)+\frac{1}{5}\right]-\frac{1}{5}\right) - \frac{9}{2}w(\al)^2 = 0\,,
\ee
where
\be \label{De}
  \De(\al)=\frac{\lb2Q\rho-\al,\al\rb}{2}\,,
\ee
and 
\be \label{w}
  w(\al)=i\sqrt{\frac{48}{22+5c}}\; \lb\al-Q\rho,h_1\rb\lb \al-Q\rho,h_2\rb\lb \al-Q\rho,h_3) \,.
\ee
In (\ref{w}) the $h_i$ are the weights of the fundamental representation of the $A_2$ Lie algebra, \textit{cf.}~(\ref{hs}). If we write $\al=\al^1\La_1 +\al^2 \La_2$, then it turns out that there are six one parameter solutions to (\ref{Dew}):
\bea \label{sings}
&&\al = \vka \La_1 \,,\qquad \al = \vka \La_1 +2Q\La_2\,, \qquad\al = \vka \La_2\,, \qquad\al = 2Q\La_1+\vka \La_2\,, 
\non \\
 &&\al = \vka \La_1 - (\vka - Q)\La_2\,,\qquad \al = \vka \La_1 - (\vka - 3Q)\La_2
\eea
Using that in our conventions $\vep=-Q$, we see that both the conditions (\ref{spec}) as well as the condition used in \cite{Warnaar:2009} belong to the set (\ref{sings}). In addition to these three solutions there are three more which differ from the other ones only when $Q\neq0$. It is known that in the $A_2$ Toda theory these six possibilities do not correspond to distinct states,  rather they are related via the so called shifted Weyl group acting on the momenta which changes the corresponding vertex operators by the so called reflection amplitudes\footnote{We thank Yuji Tachikawa for clarifying this point.}. Therefore there should also be a simple relation between the corresponding matrix integrals; in particular, if one can be explicitly evaluated then that should also be the case for the others.

\subsection{Higher--point functions}

As in the $A_1$ case we can analyse a certain class of four--point correlation functions exactly. We need the following result \cite{Warnaar:2007}
\bea \label{ekAr}
&&\int \prod_{iI} \D \la^I_i \,\, e_\ell(\la_1) \!\! \prod_{(i,I)<(j,J)} |\la^I_i-\la_j^J|^{\beta A_{ij}}\prod_{i,I} (\la^I_i)^{ \al^i_1/\vep_1} (1-\la^I_i)^{ \al^i_2/\vep_1} \\
&=&  \!\!\!S_r(\al_1,\al_2,\bet)  \binom{N_1}{\ell} \!\prod_{i=1}^r \prod_{I=1}^\ell \frac{[\al_2^1+\cdots +\al_2^i]/\vep_1+i+(N_1-I-i+1)\bet}{[\al_2^1{+}\cdots {+}\al_2^i+\al_1^i]/\vep+i+1{+}(N_1{+}N_i{-}N_{i+1}{-}I{-}i)\bet}, \non
\eea
where the $A_r$ Selberg integral $S_r(\al_1,\al_2,\bet)$ was defined in (\ref{SelAr}),   $\al_1^i = \vka \de_{i r}$, and  $e_\ell(\la_1)$ is the $\ell$th elementary symmetric polynomial defined as
\be
e_\ell(\la_1) = \sum_{I_1<\cdots <I_\ell} \la_1^{I_1} \cdots \la_1^{I_\ell}\,.
\ee
From the result (\ref{ekAr}) it follows that 
\bea
&&\!\!\!\!\!\!\!\! S_r(\al_1,\al_2,\bet; z) = \int \prod_{iI} \D \la^I_i  \prod_{(i,I)<(j,J)} |\la^I_i-\la_j^J|^{\beta A_{ij}}\prod_{i,I} (\la^I_i)^{ \frac{\al^i_1}{\vep_1}} (1-\la^I_i)^{ \frac{\al^i_2}{\vep_1}} \prod_{I} (z - \la_1^I) 
\non \\
&=&  S_r(\al_1,\al_2,\bet) \sum_{\ell =0}^{N_1} z^\ell (-1)^{N_1-\ell} \binom{N_1}{\ell} \non \\ 
&&\times \prod_{i=1}^r \prod_{I=1}^{N_1-\ell} \frac{[\al_2^1+\cdots +\al_2^i]/\vep_1+i+(N_1-I-i+1)\bet}{[\al_2^1+\cdots +\al_2^i+\al_1^i]/\vep_1+i+1+(N_1+N_i-N_{i+1}-I-i)\bet} 
\non  \\
&=& S_r(\al_1,\al_2,\bet ; 0) \sum_{\ell =0}^{N_1} \frac{z^\ell}{\ell!} (-1)^{\ell} \frac{N_1!}{(N_1-\ell)!} \\ 
&&\times \prod_{i=1}^r \prod_{I=0}^{\ell-1} \frac{-\frac{1}{\vep_2}(\al_2^1+\cdots +\al_2^i+\al_1^i+i\vep_1+\vep_1 ) + I+N_i-N_{i+1}-i }{-\frac{1}{\vep_2}( \al_2^1+\cdots +\al_2^i+i\vep_1) + I-i+1 } \non\,.
\eea
As above one can show using various Lie algebra results (see appendix \ref{Alie} for further details) that 
\be
-\vep_2(N_i - N_{i+1}) = -A_{i1}^{-1}\vep_1+A_{i+1,1}^{-1}\vep_1  - A_{ir}^{-1}\vka +A_{i+1,r}^{-1}\vka+ \lb  \al_2, h_{i+1} \rb  -  \lb  \al_0, h_{i+1} \rb \,.
\ee
 where we have used  $\al_1=\vka \La_r$ and $\al_3=\vep_1 \La_1$. Furthermore,
\be
-A_{i1}^{-1}+A_{i+1,1}^{-1} = -\frac{1}{r+1} \,,\qquad \al_1^i  - A_{ir}^{-1}\vka +A_{i+1,r}^{-1}\vka = \frac{\vka}{r+1}\,.
\ee
Using these results together with 
\be
\al_2^1+\cdots+\al_2^i =  - \lb  \al_2, h_{i+1} - h_1 \rb \,,\qquad
i = -\lb \rho, h_{i+1}-h_1 \rb\,,
\ee
it follows that
\bea \label{Ar4pt}
S_r(\al_1,\al_2,\bet; z) &=& S_r(\al_1,\al_2,\bet ; 0)  \sum_{\ell =0}^{N_1}  \frac{z^\ell}{\ell!} \frac{(A_1)_{\ell} \cdots (A_{r+1})_{\ell}}{(B_1)_{\ell} \cdots (B_{r})_{\ell} } \non \\
&=& S_r(\al_1,\al_2,\bet ; 0) {}_{r+1}F_{r}(A_1,\ldots,A_{r+1};B_1,\ldots,B_r;z)
\eea
where $(X)_n= X (X+1) \cdots (X+n-1)$ is the Pochhammer symbol,
\be \label{ArB}
B_i = -\frac{1}{\vep_2} ( \al_2^1+\cdots +\al_2^i+i\vep_1) -i + 1 = \frac{1}{\vep_2} \lb \al_2 + \vep \rho ,h_{i+1}-h_1\rb + 1 
\ee
and
\be \label{ArA}
A_i = -\frac{1}{\vep_2} \left( \frac{\vka}{r+1} + \vep_1 \frac{r}{r+1}+\lb \al_2 + \vep \rho,h_1\rb - \lb \al_0 + \vep \rho,h_{i}\rb  \right)  \,.
\ee
Note that  $(-1)^\ell \frac{N_1!}{(N_1-\ell)!}=(A_1)_{\ell}$ and that the sum over $\ell$ in (\ref{Ar4pt}) can be extended to $\infty$ since $(A_{1})_{\ell} = 0$ for $\ell > N_1$.

To compare the result (\ref{Ar4pt}) with the corresponding  result in the $A_r$ Toda theory we recall that in \cite{Fateev:2005,Fateev:2007b} it was shown that the correlator
\be
 \lb V_{-b\La_1}(z) V_{\al_1}(0) V_{\al_2} (\infty) V_{\vka\La_r}(1) \rb 
\ee
satisfies a differential equation of hypergeometric type whose solutions involves ${}_{r+1}F_{r}(A_1,\ldots,A_{r+1};B_1,\ldots,B_r;z)$ where 
\bea \label{ArAB}
A_i &=& b \left( \frac{\vka}{r+1} -b \frac{r}{r+1} +\lb \al_1-Q\rho,h_1\rb +\lb\al_2-Q\rho,h_{i}\rb \right) \non  \\
B_i &=& 1 + b\lb\al_1-Q\rho,h_1-h_{i+1}\rb 
\eea
Replacing $\al_1\rar\al_2$,  $\al_2-Q\rho\rar-(\al_0-Q\rho)$ and using the rule (\ref{epb}) 
we see that (\ref{ArAB}) agrees perfectly with (\ref{ArB}) and (\ref{ArA}).

One can also show that the Nekrasov partition function leads to the same result, see~\cite{Mironov:2009c} for a discussion. It should also be possible to analyse the correlation functions involving the insertion of $V_{ -\La_1/b}$ as well as the case with several insertions of $V_{-b \La_1}$. 

To analyse general correlation functions without imposing the above restrictions on the momenta is much more difficult. However, we stress that matrix model perturbation theory can handle any correlation function, although this method appears to be somewhat cumbersome and it is not clear what the meaning of the resulting expansion is in the $2d$ CFT.

\subsection{The curve}

We now turn to the discussion of the loop equations and the large $N$ matrix model curve. In \cite{Dijkgraaf:2009} and in sections \ref{A1curve1} and \ref{A1curve2} some examples of curves in the $A_1$ case were presented. Here we mainly focus on the $A_2$ case. We call the two matrices $\Phi$ and $\tilde{\Phi}$ and the associated potentials $W$ and $\tilde{W}$. The (non--hyperelliptic) curve is known to be of the form \cite{Kharchev:1992,Kostov:1992}
\be
x^3 = r(z) \, x + s(z)\,,
\ee
where 
\bea
r(z) &=& \frac{1}{3}[W'(z)^2+\tilde{W}'(z)^2+W'(z)\tilde{W}'(z)]  \\ 
&&-g_s\left< \tr\left(\frac{W'(z)-W'(\Phi)}{z-\Phi} \right)\right>-g_s\left< \tr\left(\frac{\tilde{W}'(z)-\tilde{W}'(\tilde{\Phi})}{z-\tilde{\Phi}} \right)\right> \non \,,
\eea
and \cite{Klemm:2003}
\bea \label{s}
s(z)& =& -\frac{1}{27}[W'(z)+2\tilde{W}'(z)] [2W'(z)+\tilde{W}'(z)] [W'(z)-\tilde{W}'(z)]   \non \\ &&
+g_s \tilde{\om}_r(z) \left< \tr\left(\frac{W'(z)-W'(\Phi)}{z-\Phi} \right)\right>-g_s\om_r(z) \left< \tr\left(\frac{\tilde{W}'(z)-\tilde{W}'(\tilde{\Phi})}{z-\tilde{\Phi}} \right)\right>  \non \\&&
-g_s^2\left< \tr\left[\frac{\D}{\D\Phi}\left(\frac{W'(z)-W'(\Phi)}{z-\Phi} \right)\right]\right>+g_s^2\left< \tr\left[\frac{\D}{\D\Phi}\left(\frac{\tilde{W}'(z)-\tilde{W}'(\tilde{\Phi})}{z-\tilde{\Phi}} \right)\right]\right>\non \\&&
+g_s\left< \tr\left(\frac{W'(z)-W'(\Phi)}{z-\Phi} \right)\right>-g_s\left< \tr\left(\frac{\tilde{W}'(z)-\tilde{W}'(\tilde{\Phi})}{z-\tilde{\Phi}} \right)\right>.
\eea
where 
\be
\om_r(z) = \frac{1}{3}(2W'(z)+\tilde{W}'(z))\,,\qquad \tilde{\om}_r(z) = \frac{1}{3}(W'(z)+2\tilde{W}'(z))\,.
\ee
(The expressions on the last two lines in (\ref{s}) can be simplified in the eigenvalue basis by using the saddle--point equations, but we will not need the resulting expression  here.)

Inserting the explicit expressions for $W$ and $\tilde{W}$ as sums of logarithms\footnote{Since the curve is derived in the limit of vanishing $\vep$ the AGT relation is $\al_i=m_i$ using a suitable definition of the masses.}
\be
W(\Phi) = \sum_{i=1}^{p} m_i \log(z_i - \Phi) \,, \qquad \tilde{W}(\tilde{\Phi}) = \sum_{i=1}^{p} \tilde{m}_i \log(z_i - \tilde{\Phi}) \,,
\ee
and using the above expressions for $r(z)$ and $s(z)$ we find
\be
x^3 = \frac{P^{(2)}_{2p-2}(z)}{ \prod_{i=1}^{p} (z-z_i)^2 } \, x + \frac{P^{(3)}_{3p-3}(z)}{ \prod_{i=1}^{p} (z-z_i)^3}\,,
\ee
where $P_{2p-2}^{(2)}(z)$ and $P_{3p-3}^{(3)}(z)$ are polynomials of degree $2p-2$ and $3p-3$, respectively. This curve is of precisely the right form to agree with the expression in \cite{Gaiotto:2009} (which was obtained by starting from the earlier result \cite{Witten:1997}). However, there is a further property that should to be checked. 

Recall that, in the $A_2$ case, there are two types of punctures, one full and one basic. For the basic (or special) puncture there is a relation between $m_i$ and $\tilde{m}_i$. In our conventions that relation is $\tilde{m}_i=0$.  Now for a  special puncture at $z=z_i$ there is a relation  between $P^{(2)}(z)$ and $P^{(3)}(z)$ that has to be satisfied at that location, see (3.25) in \cite{Gaiotto:2009}. To check that this condition holds for the matrix model curve it is sufficient to focus on a single special puncture which we can take to be located at $z_i=0$, \textit{i.e.}, $W(\Phi)= m \log(\Phi)$ and $\tilde{W}(\tilde{\Phi})= 0$. We need to check that $4 r(z)^3 - 27 s(z)^2$ scales like $\frac{1}{z^4}$ (naively it would scale like $\frac{1}{z^6}$). In other words we need
\bea \label{cond1}
&&4 ( \frac{1}{3}[W'(z)^2+\tilde{W}'(z)^2+W'(z)\tilde{W}'(z)] )^3  \\
&-&\!\!\! 27( -\frac{1}{27}[W'(z)+2\tilde{W}'(z)] [2W'(z)+\tilde{W}'(z)] [W'(z)-\tilde{W}'(z)] )^2=0 ,\non
\eea
and
\bea \label{cond2}
&& 12 ( \frac{1}{3}[W'(z)^2+\tilde{W}'(z)^2+W'(z)\tilde{W}'(z)] )^2(f(z)+ \tilde{f}(z)  ) \\ 
&+&\!\!\!2[W'(z){+}2\tilde{W}'(z)] [2W'(z){+}\tilde{W}'(z)] [W'(z){-}\tilde{W}'(z)] (\om_r(z)  \tilde{f}(z) -\tilde{\om}_r(z) f(z) ) =0, \non 
\eea
where
\be
f(z) = -g_s\left< \tr\left(\frac{W'(z)-W'(\Phi)}{z-\Phi} \right)\right> , \,\tilde{f}(z) = -g_s\left< \tr\left(\frac{\tilde{W}'(z)-\tilde{W}'(\tilde{\Phi})}{z-\tilde{\Phi}} \right)\right> .
\ee
\noindent
Both the above equations are easily shown to hold for the special puncture with $\tilde{m}_i=0$ thereby establishing the equivalence with the results in \cite{Gaiotto:2009}\footnote{There is also another solution to (\ref{cond1}) and (\ref{cond2}), \textit{viz.}~$m_i=\tilde{m}_i$. This solution is precisely the alternative solution discussed at the end of section \ref{Arthree} (note that the curve is derived for $\vep=0$). }. 

For higher rank curves the general structure of the matrix model curve is also known; see \cite{Chiantese:2003} for the state--of--the--art knowledge. The matrix model curves can be analyzed and compared to the gauge theory curve as above.

\setcounter{equation}{0} 
\section{$5d$ gauge theories and $q$--deformed matrix models} \label{sq}

Nekrasov partition functions can also be defined for supersymmetric gauge theories in five dimensions formulated on $\RR^4\times S^1$ \cite{Nekrasov:2002}. As an example, in the $\SU(2)$ theory with four matter hypermultiplets in the fundamental representation the instanton partition function is 
\be \label{5dzinst}
\mathcal{Z}_{\rm inst}=\sum_{\vec{Y}} y^{|\vec{Y}|}  \prod_{m,n=1}^{2} \prod_{s\in Y_m} \frac{ \mathcal{P}(\ha_m,Y_m,s) )}{\mathcal{E}(\ha_m-\ha_n,Y_m,Y_n,s)(\mathcal{E}(\ha_m-\ha_n,Y_m,Y_n,s) -\vep) }
\ee
where $(\ha_1,\ha_2)=(a,-a)$ and
\bea
\mathcal{E}(x,Y_m,Y_n,s) &=& \sinh(R(x-\ep_1 L_{Y_n}(s) + \ep_2(A_{Y_m}(s)+1) )) \,, \non \\
\mathcal{P}(\ha_m,Y_i,s) &=& \prod_{f=1}^{4} \sinh(R(\ha_m-(j-1)\ep_1-(i-1)\ep_2 - m_f))
\eea
with $R$ the radius of the $S^1$. See section \ref{squiver} for more details about the notation. 
The expression (\ref{5dzinst}) can also be obtained from topological string considerations see \textit{e.g.}~\cite{Iqbal:2003} \cite{Iqbal:2007}). The partition function also has a perturbative piece whose explicit expression can be found \textit{e.g.}~in \cite{Nekrasov:2002,Nakajima:2005}.

The question we would like to address in this section is: Are there matrix model and CFT descriptions of partition functions of the type (\ref{5dzinst})?

In the recent paper \cite{Awata:2009} a proposal was made for the CFT description of the pure $\SU(2)$ theory in five dimensions. This proposal involved the so called $q$--deformed Virasoro algebra \cite{Shiraishi:1995}. It is very natural to expect that there is  a CFT description of $5d$ (conformal) quiver gauge theories which involves $q$--deformed Virasoro  \cite{Shiraishi:1995} and $q$--deformed $\cW$ algebras \cite{Feigin:1995}.  Unfortunately the representation theory of these algebras is not very well developed. Analogues of the primary fields and their quantum numbers are known, but the analogue of \textit{e.g.}~the relation
\bea
[L_m, V_{\al}] &=&  z^m[ \,(m+1) \, \De(\al) \, V_\al + z\, (L_{-1} V_\al) \,] \non\\
&=&  z^m( \,m\, \De(\al)\,  V_\al +  [L_0,V_\al]  \,) \,,
\eea  
is not known (as far as we know). This fact complicates the analysis and makes direct calculations of chiral blocks difficult.  

Instead, we try to obtain a matrix model description. A natural starting point is to look for a generalisation of the $A_1$ three--point function (\ref{A1mm3}).

There is a known $q$--deformation of (\ref{A1mm3}) in the literature, which can be written 
\be
\int \prod_{I=1}^N \D_q \la^I \, \prod_{I=1}^N (\la_I)^{2\al_1/\vep_1} \frac{(q^{-\al_2/\vep_1} \la^I;q)_{\infty}}{(q^{\al_2/\vep_1} \la_I;q)_{\infty}} \prod_{I<J} (\la^J)^{2\bet}  \frac{(q^{-\bet} \la^I/\la^J;q)_{\infty}}{(q^\bet \la^I/\la^J)_{\infty}}  \,,
\ee
where $0<q<1$ and $\int \D x_q$ is the so called $q$-integral (or Jackson integral) defined via
\be
\int_0^1 \D_q x f(x) = (1-q) \sum_{k=0}^{\infty} f(q^k) q^k\,.
\ee
In the limit $q\rar 1^-$ this expression converges to the Riemann integral $\int_{0}^{1} f(x) \D x$. Furthermore, $(a;q)_{\infty}= \prod_{k=0}^{\infty} (1-a q^k)$, and we also use the notation:
\be
(a;q)_{\ell} = (1-a)(1-a q)\cdots(1-aq^{\ell-1}) = \frac{ (a ;q)_{\infty} }{ (q^{\ell} a;q)_{\infty} }\,.
\ee

Based on the above expression we tentatively propose the rule
\be
\hat{V}_\al(z) \equiv \prod_I (z-\la^I)^{2\al/\vep_1} \rar \prod_I z^{2\al/\vep_1} \frac{( q^{-\al/\vep_1}\la^I/z;q)_{\infty} }{ (q^{\al/\vep_1}\la^I/z ,q )_{\infty} } \equiv \hat{V}^q_\al(z)\,.
\ee
In the limit $q\rar 1^-$, $ \hat{V}^q_\al(z) \rar \hat{V}_\al(z)$. 
In the special case $z=0$ we assume that  $ \hat{V}^q_\al(0)=  \hat{V}_\al(0) = \prod_I (\la^I)^{2\al/\vep_1}$. Note that when $\al=\vep_1/2$,  $V^q_{\al}(z) = \prod_I z \frac{( q^{-1/2}\la^I/z;q)_{\infty} }{ (q^{1/2}\la^I/z ,q )_{\infty} }$  reduces to $\prod_I (z-q^{-1/2}\la^I) $.

The above three--point function can be evaluated (at least when $\bet$ is an integer) and leads to a product of $q$--gamma functions, but since we are not aware of a $q$--analogue of (\ref{Gaid}) we will not discuss the result here. 

Instead we turn to the four--point function with $\al_3=\vep_1/2$. Using the result in \cite{Mimachi:1998} the resulting expression  can be explicitly evaluated 
\bea \label{q4pt}
&&\!\!S^q(\al_1,\al_2,\bet;z) = \int \prod_{I=1}^N \D_q \la^I \,  (z{-}\frac{\la^I}{q^{1/2}})  (\la_I)^{\frac{2\al_1}{\vep_1}} \frac{(q^{-\frac{\al_2}{\vep_1}} \la^I;q)_{\infty}}{(q^{\frac{\al_2}{\vep_1}} \la_I;q)_{\infty}} \prod_{I<J} (\la^J)^{2\bet}  \frac{(q^{-\bet} \frac{\la^I}{\la^J};q)_{\infty}}{(q^\bet \frac{\la^I}{\la^J})_{\infty}}  \non \\
&&\!\!=\, S^q(\al_1,\al_2,\bet;0)\,  {}_2\phi_1(q^{-\bet N},q^{2\al_1/\vep_1+2\al_2/\vep_1+2) + \bet N - \bet};q^{2\al_1/\vep_1+1};q^\bet,\tilde{z} )\eea
where $\tilde{z}= z\, q^{-\frac{2\al_2/\vep_1-1}{2}}$ and 
\be
{}_2\phi_1(A,B;C;q,z) = \sum_{k=0}^{\infty} \frac{(A;q)_k (B;q)_k }{ (C;q)_k (q;q)_k } z^k \,.
\ee
(Actually, since  $N$ is an integer ${}_2\phi_1$ in the above expression reduces to a so--called little $q$--Jacobi polynomial.)

The integral (\ref{q4pt}) was also discussed in~\cite{Kaneko:1996}, albeit in a somewhat different guise. In that paper it was shown that the integral satisfies a certain difference equation. This is a $q$--analogue of the result in \cite{Kaneko:1993} (\textit{cf.}~the discussion in section \ref{A1high}). However, the analysis  in  \cite{Mimachi:1998} is more transparent, although the case corresponding  to multiple insertions of $\hat{V}^q_{\vep_1/2}$ is not discussed in  \cite{Mimachi:1998}. 

The above result (\ref{q4pt}) can be rewritten
\bea \label{5d4pt}
&& {}_2\phi_1(q^{-\bet N},q^{(2\al_1/\vep_1+2\al_2/\vep_1+2)+\bet N-\bet)};q^{(2\al_1/\vep_1+1)};q^\bet,\frac{z}{q^{2\al_2/\vep_1-1/2}} )  \\
&&=\, \sum_{\ell=0}^{\infty} \prod_{k=0}^{\ell-1} \frac{\sinh( R [- N+  k]) \sinh( R [ \bet^{-1} (\frac{2\al_1}{\vep_1}+\frac{2\al_2}{\vep_1}+2)+ N-1+ k])}{\sinh(R [\bet^{-1}(2\al_1/\vep_1+1)+ k]) \sinh(R [1+ k])   } (z q^{1-\bet})^\ell,\non
\eea
where we have used $q = \rme^{-2R/\bet}$. 

To compare this expression with the one arising from the Nekrasov partition function, we go through the same steps as in section \ref{A1high} and make the choices $m_1=a$ and $m_2 =-a+\ep_1$ which implies that (\ref{5dzinst}) reduces to 
\be
\sum_{\ell=0}^{\infty} y^\ell \prod_{k=0}^{\ell} \frac{\sinh(R[(\al_1{+}\al_2{+}\al_4{-}\frac{\ep_1}{2}] )/\ep_2+k)\sinh(R[(\al_1{+}\al_2{+}\al_4{-}\frac{3}{2}\ep_1{-}\ep_2)/\ep_2+k])}{\sinh(R[(\ep_1-2\al_1)/\ep_2+k])\sinh(R[1+k])},
\ee 
where we have also used the AGT relation (\ref{mal}).
This expression is readily seen to agree with (\ref{5d4pt}) using $y=z q^{1-\bet}$ and the same arguments as in section (\ref{A1high}), \textit{cf.} the discussion after eq.~(\ref{ABC4}). 

In \cite{Kaneko:1996} there is also an extension of the above result to the case with multiple insertions of $\hat{V}^q_{\vep_1/2}$. In this case the result involves the function ${}_2\Phi^{(q,t)}_1(A,B;C;z)$ which is a $q$ analogue of (\ref{hyperJack}) and involves Macdonald polynomials rather than Jack polynomials, see \cite{Kaneko:1996} for further details. So far we have only considered the $A_1$ case; it should also be possible to consider the $A_r$ case using the results in \cite{Warnaar:2007}.
 
We close this section with a few words of caution. There are in general several possible $q$--deformations and the one above may not be the right one. Also, we should mention that in \cite{Klemm:2008} another deformation of the matrix model was shown to be related to Nekrasov partition functions for five--dimensional gauge theories. The deformation in \cite{Klemm:2008} replaces the Vandermonde determinant with $\prod_{I<J} \sinh(\la^I-\la^J)$. This possibility was also mentioned in \cite{Dijkgraaf:2009}.

\setcounter{equation}{0} 
\section{Discussion and outlook} \label{sdisc}

In this paper we have studied the $A_r$ quiver matrix models which were introduced in \cite{Dijkgraaf:2009} and argued to capture correlation functions (chiral blocks) in the $2d$ $A_r$ Toda field theories and Nekrasov partition functions (instanton partition functions) in the $4d$ $A_r$ quiver gauge theories. 

From the point of view of the matrix model the expansion in $z_i$ (the locations of the vertex operators) is somewhat awkward, but we have shown that several known results can be rederived from the matrix model; some of our checks are quite non--trivial. It would be interesting to develop the matrix model technology further, and, for instance, to clarify the choice of integration contour and to develop a perturbation theory in $z_i$. 

We also made a proposal for an extension of the matrix model to capture the Nekrasov partition function of $5d$ quiver gauge theories. This speculative proposal passed a non--trivial check, but deserves further study. 

One open problem is to extend the analysis to the other $ADE$ Lie algebras. Let us make a comment about the $D_r$ case. The matrix model curve for the $D_r$ model can be extracted from~\cite{Chiantese:2003}, eq.~(3.51) and can be seen (after some changes of notation) to be of the same general form as the curves in \cite{Tachikawa:2009}. It would be interesting to study this in more detail.

\section*{Acknowledgements}
We would like to thank Yuji Tachikawa and Cumrun Vafa for comments and correspondence.  
RS is partially supported by the Funda\c{c}\~{a}o para a Ci\^{e}ncia e a Tecnologia (Portugal).

\appendix

\setcounter{equation}{0} 
\section{Appendix}

\subsection{$A_r$ roots and weights} \label{Alie}

Here we collect some standard results for the $A_r$ Lie algebras.
The root/weight space of the $A_r$ Lie algebra can viewed as a $r$--dimensional subspace of $\RR^{r+1}$. The unit vectors of $\RR^{r+1}$ will be denoted $u_i$ ($i=1,\ldots,r+1$) and satisfy $\lb u_i,u_j\rb = \de_{ij}$. The simple roots are $e_i = u_i-u_{i+1}$ ($i=1,\ldots,r$) and the positive roots are $e_{ij} =  u_i-u_j$ (with $1\leq i<j \leq {r+1}$). The Cartan matrix is $A_{ij}=\lb e_i,e_j\rb$, and its inverse is $A^{-1}_{ij} = \frac{1}{r+1}\min(i,j)[r+1-\max(i,j)]$. The Weyl vector, $\rho$, is half the sum of the positive roots; hence $\rho = \half \sum_{i=1}^{r+1} (r- 2i + 2) u_i$. The fundamental weights, $\La_i$, are defined as
\be
\La_i = u_1+\cdots+u_i - \frac{i}{r+1}\sum_{j=1}^{r+1} u_j \,, \qquad (i=1,\ldots,r)
\ee
and  satisfy $\lb \La_i ,e_j\rb = \de_{ij}$. Note that $\sum_{i=1}^r \La_i = \rho$. Finally, the weights of the fundamental representation can be chosen as 
\be \label{hs}
h_i = u_i - \frac{1}{r+1} \sum_j u_j = \La_1 - \sum_{j=1}^{i-1} e_j\,, \qquad (i=1,\ldots,r+1)
\ee
Note that $h_1 = \La_1$ and $\sum_j h_j =0$. 

\subsection{Orthogonal polynomials and the $A_1$ three--point function} \label{ortho}

Here we present an alternative evaluation method for the integral (\ref{A1mm3}) in the case $\bet=1$ using orthogonal polynomials. 
Consider the one--matrix model partition function 
\be
Z = \frac{1}{(2\pi)^NN!} \int \prod_{I=1}^N \rmd \lambda^I \,  \prod_{I<J} (\lambda^I-\lambda^J)^2 \, \rme^{\frac{1}{g_s} \sum_{i=1}^N W(\lambda^I)}\,,
\ee
\noindent
and introduce orthogonal polynomials, $\{ p_n (z) \}$, with respect to the measure
\be
\rmd \mu (z) = \rme^{\frac{1}{g_s} W(z)}\, {\rmd z}
\ee
\noindent
 as
\be\label{op}
\int_{\BR} \rmd \mu(z)\, p_n(z) p_m(z) = h_n \delta_{nm}\,, \qquad n \ge 0\,,
\ee
\noindent
where one further normalizes $p_n (z)$ such that $p_n (z) = z^n + \cdots$. Noticing that the Vandermonde determinant $\prod_{I<J} (\la^I-\la^J)^2$ equals $\det p_{J-1} (\lambda^I)$, the one--matrix model partition function may be computed as 
\be\label{zop}
Z = \frac{1}{(2\pi)^N}  \prod_{n=0}^{N-1} h_n \,.
\ee
\noindent 

In the case of interest to us, the potential is
\be \label{penpot}
W(z) = \tr\, \sum_{a=1}^k 2\al_a \log(z_a - z)\,,
\ee
\noindent
which, in principle, forbids the use of standard orthogonal polynomial techniques. However, the fact that the non--polynomial structure is logarithmic actually allows us to get around this issue when $k=2$, as we shall see now.
Indeed, in this case (setting $z_1=0$ and $z_2=1$) the measure associated with (\ref{penpot}) becomes 
\be
\rmd \mu(z) = \left(1- z\right)^{2\al_2} z^{2\al_1}\, {\rmd z}\,,
\ee
\noindent
and is immediately related to the orthogonal polynomial family of Jacobi polynomials. 
The combination
\be
J^{(\alpha,\gamma)}_n (z) \equiv \frac{ n!\, \Gamma \left( n+\alpha+\gamma+1 \right)}{\Gamma \left( 2n+\alpha+\gamma+1 \right)}\, P^{(\alpha,\gamma)}_n (2z-1) \,,
\ee
\noindent
where $P^{(\alpha,\gamma)}_n (z)$ is a Jacobi polynomial, is normalized such that $J^{(\alpha,\gamma)}_n (z) = z^n + \cdots$ and satisfies 
\be
\int_{0}^{1} {\rmd z} \left( 1-z \right)^\alpha z^\gamma J^{(\alpha,\gamma)}_n (z)\, J^{(\alpha,\gamma)}_m (z) = h_n \delta_{nm}\,,
\ee
with
\be
h_n = n! \frac{\Gamma \left( n+\alpha+1 \right) \Gamma \left( n+\gamma+1 \right) \Gamma \left( n+\alpha+\gamma+1 \right)}{\Gamma \left( 2n+\alpha+\gamma+2 \right) \Gamma \left( 2n+\alpha+\gamma+1 \right)}.
\ee
\noindent
Using (\ref{zop}) this immediately leads to the \textit{exact}  result
\bea
Z &=& \frac{1}{(2\pi)^N} \prod_{n=0}^{N-1}  n! \frac{\Gamma \left( n+2\alpha_2/g_s+1 \right) \Gamma \left( n+\alpha_1/g_s+1 \right) \Gamma \left( n+\alpha_2+\alpha_1+1 \right)}{\Gamma \left( 2n+\alpha_2+\alpha_1+2 \right) \Gamma \left( 2n+\alpha_2+\alpha_1+1 \right)} \non \\
&=& \frac{1}{(2\pi)^N} \prod_{n=0}^{N-1}  n! \frac{\Gamma \left( n+\alpha_2+1 \right) \Gamma \left( n+\alpha_1+1 \right) }{\Gamma \left(N + n+1 +\alpha_2+\alpha_1\right)}\,.
\eea
which agrees with (\ref{A1sel}) (when $\bet=1$).

\begingroup\raggedright\endgroup

\end{document}